\documentclass[a4paper,12pt]{article}

\usepackage{authblk}
\usepackage{array}
\usepackage{hyperref}
\hypersetup{colorlinks=true, linkcolor=blue, citecolor=blue, urlcolor=blue}
\usepackage{braket}
\usepackage{amssymb}
\usepackage{amsmath}
\usepackage{amscd}
\usepackage[all,arc,knot,matrix]{xy}
\usepackage{graphicx}
\usepackage{tikz}
\usetikzlibrary{angles,quotes}
\usepackage{geometry}
\numberwithin{equation}{section}
\newcommand{\dd}{\mathrm{d}}

\begin{document}

\title{Regge calculus for $f(R)$ gravity with torsion}
\author[1]{Xiao-Kan Guo\footnote{E-mail: kankuohsiao@whu.edu.cn}}
\affil[1]{School of Mathematics \& Physics, Yancheng Institute of
Technology, Yancheng 224051,  Jiangsu, China}
\date{\today}
\maketitle

\begin{abstract}
We formulate $f(R)$ gravity with spacetime torsion in Regge calculus, but
without an \emph{ad hoc} geometric definition of torsion. Working in
D'Adda's formulation of simplicial gravity with coordinates, we show that
the axial sector of the torsion is carried by a massless Kalb--Ramond
two-form, which in four dimensions is dual to a single scalar on the dual
vertices. The discretized theory is therefore pure Regge gravity coupled to
two scalar fields, the scalaron and the axion, without a geometric
definition of torsion on the Regge lattice. We derive the lattice action
and the complete set of discrete equations of motion, and we verify the
continuum limit. For a homogeneous and isotropic lattice we obtain the
discrete Friedmann--Raychaudhuri system, compute the exact kinetic and
curvature coefficients for a regular triangulation of $\mathbb S^3$, and
show that the discrete equations reproduce the continuum closed solution at
first order in the evolution time step.  The results therefore present a very special case in which the torsional effects are completely described by scalar fields on the Regge lattice and there is no need of a full discrete differential calculus.
\end{abstract}

\section{Introduction}\label{sec:intro}
Regge calculus \cite{Regge1961,ReggeWilliams2000} encodes the geometry of a
$d$-dimensional triangulated
spacetime in the edge lengths, with curvature concentrated on the
$(d-2)$-dimensional hinges as deficit angles. It is general relativity
without coordinates, and for this reason it is the natural lattice
formulation of the metric sector of gravity. But the same feature makes the
coupling to torsion problematic. Torsion is the antisymmetric part of the
affine connection. Although it is sourced by the spin density of matter
\cite{Hehl1976,Shapiro2002},  in metric-affine $f(R)$ theories it belongs
to the gravitational sector itself. A Regge lattice, however, carries no
local frame, so there is no natural place to store the torsion. 

There have been many attempts at defining a geometric torsion on a Regge lattice \cite{Drummond1986,SchmidtKohler2001,PereiraVargas2002,Xue2009,Xue2010,Xue2012,Yan2026,Kisielowski2019,Christiansen2023,DAdda2021}. For instance, one may represent
torsion by geometric defects, either as dislocations carrying a Burgers vector
\cite{SchmidtKohler2001,PereiraVargas2002} or as a constant body torsion inside
each simplex \cite{Drummond1986}. This requires a geometric definition of
torsion, and the relation between the Burgers vector and the torsion tensor
becomes ambiguous when the curvature does not vanish \cite{Yan2026}. In the
Einstein-Cartan lattice, the Burgers vector vanishes in the absence of matter
\cite{SchmidtKohler2001}, whereas in the teleparallel lattice it does not
\cite{PereiraVargas2002}. A second route keeps a tetrad and a spin connection
as independent variables and applies the Cartan structure equation on the
triangles \cite{Xue2009,Xue2010,Xue2012,Yan2026}. It is geometrically
transparent but retains all 24 torsion components and a local frame on every
simplex, which is what the Regge formulation is designed to avoid. A third
route uses discrete differential forms \cite{DAdda2021}. Here the difficulty
is structural. Forms placed on dual cells violate the Leibniz rule, while
forms placed on the simplices satisfy it but lose associativity, so a
consistent discrete exterior calculus for $p\ge2$ forms is not available.
Each route therefore meets an obstruction.

The way out is to count the independent components of the torsion. In four
dimensions, the contorsion tensor has 24 components. In the metric-affine $f(R)$ gravity
the contorsion is non-dynamical. The field equations fix 20 of the 24
components algebraically. The vector trace is slaved to the scalaron, and the
tensor part vanishes on shell. The remaining four components form the
axial (totally antisymmetric) part, which is carried by a massless
Kalb--Ramond two-form \cite{Kumar2024,Kalb1974,Scherk1974}. A massless two-form
carries a single degree of freedom and is dual to a scalar. A scalar, unlike a
two-form, lives on the vertices of the dual lattice. The torsion sector can
therefore be carried by a single scalar on the dual vertices, without the need
for $p\ge2$ forms or a geometric definition of torsion. The purpose of this
paper is to  work out its consequences on
the lattice.

From the work of Kumar {\it et al.} \cite{Kumar2024} we know that the torsion sector of metric-affine $f(R)$ gravity in the Einstein frame is exactly
equivalent, at the classical level, to a single massless axion $a$ with a
field-dependent kinetic term. By working with  D'Adda's formulation of
simplicial gravity with coordinates \cite{DAdda2021},  this theory discretizes to
a pure Regge action coupled to two vertex scalars,
the scalaron $\phi(\alpha)$ and the axion  $a(\alpha)$.
The lattice action yields a complete set of discrete equations of motion,
namely the axion current conservation equation, the scalaron
Klein--Gordon equation, and the edge equation in which the deficit-angle
variation drops out by the Sch\"afli identity. The
edge equation reduces in the continuum limit to the Einstein equation of the
Einstein-frame theory. The axial torsion vector is the gradient of the axion,
$A_\mu=\partial_\mu a$, so torsion-free configurations correspond to a
constant axion. The torsion of the present formulation is therefore the axial
projection of the torsion of the first-order and holonomy formulations
\cite{Xue2009,Yan2026}. In this sense, this work is  a very special route to the Regge Calculus with torsion, which should be taken as a proof of principle, rather than a complete resolution.

Relative to Ref.~\cite{Kumar2024}, the new content of this paper is the
lattice formulation and its consequences. Ref.~\cite{Kumar2024}
establishes the continuum duality between the torsion sector and the axion.
Here we show that the axial projection is the only independent torsion
component, that D'Adda's simplicial calculus realizes the two-form as a
vertex scalar, and that the discrete equations close on the Regge lattice.
What distinguishes the construction is that it needs
neither a geometric torsion nor a discrete exterior calculus for $p\ge2$
forms.

On a homogeneous and isotropic lattice the equations of motion reduce to a
discrete Friedmann--Raychaudhuri system, extending the Regge cosmology
initiated by Collins and Williams \cite{Collins1973}. For a regular
triangulation of $\mathbb S^3$, we compute the kinetic
and curvature coefficients, and the prism triangulation
is shown to suffer from an obstruction to homogeneity. 

Interestingly, for $d\ge3$ a generic local conformal
transformation maps a piecewise-flat metric out of the Regge class and the
deficit angles are not conformally invariant.
Consequently an exact conformal duality cannot hold
between Regge-type lattice theories and can be realized only in the continuum
limit.

The paper is organized as follows. In Sec.~\ref{sec:regge} we fix the
simplicial framework, collect the identities used below, and describe
conformal transformations on a Regge metric. Section~\ref{sec:duality}
establishes the axial-torsion-axion duality and compares our torsion with
the existing proposals. Section~\ref{sec:lattice} constructs the discrete
theory and derives its equations of motion. Section~\ref{sec:frw} applies
the equations to homogeneous cosmology, and
Sec.~\ref{sec:conclusion} concludes.

We use the Lorentzian signature $(-,+,+,+)$ in the continuum and the
Euclidean convention for the simplicial lattice, as is standard in Regge
calculus. The continuum metric is $g_{ab}$ and its determinant is
$g=\det g_{ab}$; a conformal transformation acts on the metric,
$g_{ab}\mapsto\Omega^2g_{ab}$. On a simplex with coordinates $x^\mu$, the same
metric is written $g_{\mu\nu}(\alpha)$. Quantities with an overbar refer to the Levi-Civita connection and
quantities with a tilde to the Einstein frame. We set $\kappa^2=8\pi G$ and
use $\dd$ for the exterior derivative. For a form $\omega$,
$|\omega|^2\star1=\omega\wedge\star\omega$ denotes the component norm. The
antisymmetrized derivative $\partial_{[a}Z_{bc]}$ is defined as the sum over
cyclic permutations, $\partial_{[a}Z_{bc]}=\partial_aZ_{bc}+
\partial_bZ_{ca}+\partial_cZ_{ab}$, so that it equals the components of
$\dd Z$. Numerical factors arising from this convention are absorbed in the
normalizations of $B$ and $Z$.
\section{Simplicial gravity with coordinates}\label{sec:regge}

In this section we fix the framework of Regge calculus and D'Adda's simplicial gravity with coordinates \cite{DAdda2021}, and collect the elementary identities used in the rest of the paper. 

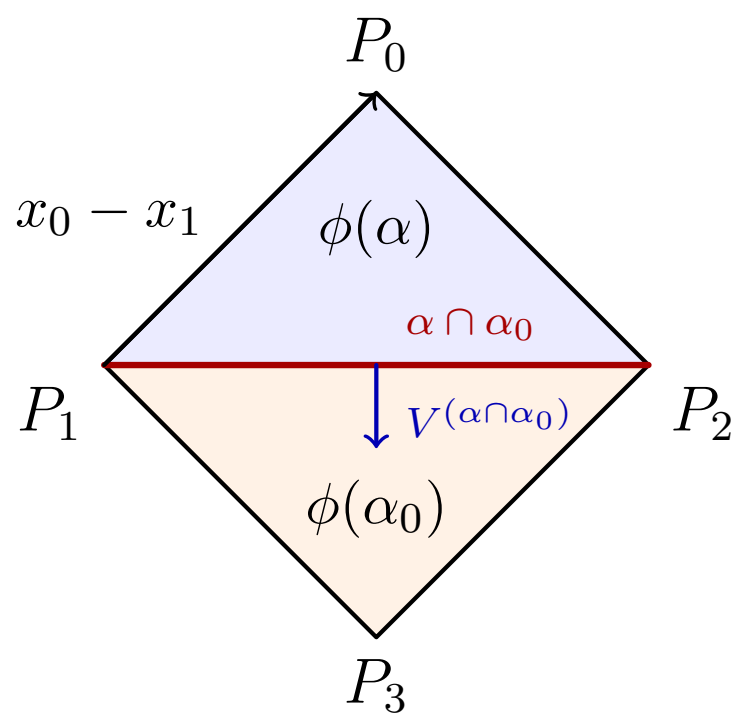
\begin{figure}[t]
\centering
\begin{tikzpicture}[scale=1.0,line join=round,line cap=round]
  \coordinate (P0) at (0,1.8);
  \coordinate (P1) at (-1.8,0);
  \coordinate (P2) at (1.8,0);
  \coordinate (P3) at (0,-1.8);
  \fill[blue!8] (P0)--(P1)--(P2)--cycle;
  \fill[orange!10] (P1)--(P3)--(P2)--cycle;
  \draw[thick] (P0)--(P1)--(P2)--cycle;
  \draw[thick] (P1)--(P3)--(P2);
  \draw[very thick,red!65!black] (P1)--(P2);
  \node[above] at (P0) {$P_0$};
  \node[below left] at (P1) {$P_1$};
  \node[below right] at (P2) {$P_2$};
  \node[below] at (P3) {$P_3$};
  \node[font=\small] at (0,0.9) {$\phi(\alpha)$};
  \node[font=\small] at (0,-0.95) {$\phi(\alpha_0)$};
  \node[above right,red!65!black,font=\scriptsize] at (0.05,0.02) {$\alpha\cap\alpha_0$};
  \draw[->,thick,blue!70!black] (0,0.0)--(0,-0.55);
  \node[right,font=\scriptsize,blue!70!black] at (0.05,-0.35) {$V^{(\alpha\cap\alpha_0)}$};
  \draw[->,thick] (P1)--(P0);
  \node[left,font=\small] at (-1.0,1.0) {$x_0-x_1$};
\end{tikzpicture}
\caption{Two-dimensional illustration of D'Adda's framework. Coordinates
$x_i$ are assigned to the vertices $P_i$, while the metric and the scalar
fields $\phi$ live on the simplices $\alpha$. The lattice derivative at
$\alpha$ uses the neighbouring simplex $\alpha_0$ across the face opposite
$P_0$, weighted by the face volume vector $V^{(\alpha\cap\alpha_0)}$. The
edge lengths are $l_{ij}^2=g_{\mu\nu}(\alpha)(x_i-x_j)^\mu(x_i-x_j)^\nu$.}
\label{fig:dadda}
\end{figure}

The  D'Adda framework
assigns coordinates $x^\mu_i$ to the vertices $P_i$ of each simplex and a
metric $g_{\mu\nu}(\alpha)$ to each simplex $\alpha$ (aka. the Regge metric). The edge lengths
$l_{ij}^2=g_{\mu\nu}(\alpha)(x_i-x_j)^\mu(x_i-x_j)^\nu$ are then computed (Fig.~\ref{fig:dadda})

The assignment of coordinates is due to the following reason. Edge lengths alone determine
the geometry but give no direction along which a field can be
differentiated. Without coordinates the only operation available on a scalar
is a difference between neighbouring simplices, which carries no vector
index. The vertex coordinates provide the edge vectors $x_i-x_j$, so that
the finite difference can be projected onto the edge direction and the
result is a genuine vector derivative. This is what makes the kinetic terms
and the coupling to the edge vectors possible. The price is that the
derivative is second-order accurate only for regular simplices, the case
used below.

\subsection{Regge calculus}

A $d$-dimensional spacetime is triangulated into $d$-simplices glued pairwise
along their $(d-1)$-faces. Each simplex is flat, and its
$\tfrac12d(d+1)$ edge lengths determine its geometry. Curvature is
concentrated on the hinges, on which the deficit angle is
\begin{equation}
\theta_h=2\pi-\sum_{\alpha\supset h}\theta^{(\alpha)}_h,
\label{eq:deficit}
\end{equation}
where $\theta^{(\alpha)}_h$ is the dihedral angle of $\alpha$ at $h$. For a hinge $h$, the integrated curvature over the dual cell $v(h)$ is
$\int_{v(h)}R\sqrt g=\theta_h$, while the volume of $v(h)$ is proportional to
$V(h)$, so the local value of $R$ is $\theta_hV(h)/v(h)$ and the Einstein
action becomes the Regge action 
\begin{equation}
S_R=\frac{1}{2\kappa^2}\sum_hV(h)\,\theta_h .
\label{eq:Regge}
\end{equation}

Two elementary identities will be used later. Consider a simplex $T$
with faces $F^{(r)}$ carrying inward unit normals $n^{(r)}$. Applying the
divergence theorem to a constant vector field $v$ on the flat simplex,
\begin{equation}
\int_T\nabla\cdot v=\oint_{\partial T}v\cdot n
\quad\Longrightarrow\quad
0=\sum_r V(F^{(r)})\,n^{(r)}\!\cdot v ,
\end{equation}
and since this holds for every constant $v$,
\begin{equation}
\sum_{r}V(F^{(r)})\,n^{(r)}=0 .
\label{eq:divergence}
\end{equation}
The same argument applied to a face $F$, regarded as a flat
$(d-1)$-simplex whose faces are the hinges $H^{(s)}$, gives
\begin{equation}
\sum_{s}V(H^{(s)})\,n^{(s)}=0 ,
\label{eq:divergence2}
\end{equation}
with $n^{(s)}$ the unit bivectors orthogonal to the hinges. It is precisely
this cancellation that renders Drummond's interface torsion ineffective
\cite{Drummond1986}.

\subsection{Scalar and Einstein actions}
 D'Adda's construction of the discrete curvature of a hinge replaces $\theta_h$ by $\sin\theta_h$
\cite{DAdda2021}, so that
\begin{equation}
\int \dd^dx\sqrt g\,R
=\sum_h V(h)\sin\theta_h+\mathcal O(l^2).
\label{eq:EHdisc}
\end{equation}
 In the rest of the paper the Regge action is
understood with this sine replacement.

Scalar fields are introduced through the dictionary
\begin{equation}
\int \dd^dx\sqrt g\;\to\;\sum_\alpha V(\alpha),\qquad
x\;\to\;\alpha,\qquad
\partial_\mu\;\to\;\hat\partial_\mu,
\label{eq:dictionary}
\end{equation}
Denoting by $\alpha_i$ the simplex sharing with $\alpha$ the face opposite to
$P_i$, the lattice derivative of a scalar $\phi(\alpha)$ is
\begin{equation}
\hat\partial_\mu\phi(\alpha)
=\frac12\sum_{i=1}^{d+1}\phi(\alpha_i)\,
\frac{V^{(\alpha\cap\alpha_i)}_\mu(\alpha)}{V(\alpha)} .
\label{eq:deriv}
\end{equation}
This operator is covariant and, for regular simplices, second-order accurate.
The fields $\phi(\alpha)$  are defined on the $4$-simplices
$\alpha$, that is, on the vertices of the dual complex. They are constant on
each simplex and are not attached to the primal vertices $P_i$, which carry
only the embedding coordinates. 

Under a discrete change
of coordinates $x\to x'(x)$ at the vertices, with
$g'_{\mu\nu}(\alpha)=(\partial x^\rho/\partial x'^\mu)
(\partial x^\sigma/\partial x'^\nu)g_{\rho\sigma}(\alpha)$, one has
$\hat\partial'_\mu\phi=(\partial x^\rho/\partial x'^\mu)\hat\partial_\rho\phi$.
Expanding $\phi(\alpha_i)$ about $\hat x(\alpha)$, the zeroth-order term
cancels because $\sum_iV^{(\alpha\cap\alpha_i)}_\mu=0$
[Eq.~\eqref{eq:divergence}], the first-order term reproduces the gradient,
and the remainder is $\mathcal O(l^2)$, so that
\begin{equation}
\hat\partial_\mu\phi(\alpha)=\partial_\mu\phi(\hat x(\alpha))
+\mathcal O(l^2).
\label{eq:derivcont}
\end{equation}
In the Euclidean convention used for the lattice, the action for scalar $\phi(\alpha)$ is therefore
\begin{equation}
S_\phi=\sum_\alpha V(\alpha)\Big[
\tfrac12 g^{\mu\nu}(\alpha)\hat\partial_\mu\phi(\alpha)
\hat\partial_\nu\phi(\alpha)+U(\phi(\alpha))\Big],
\label{eq:scalar}
\end{equation}
where $U$ denotes the scalar potential.
Inserting \eqref{eq:deriv}, the kinetic part becomes
\begin{equation}
S_{\phi,\rm kin}=\frac18\sum_\alpha
\frac{g^{\mu\nu}(\alpha)}{V(\alpha)}
\sum_{i,j}V^{(\alpha\cap\alpha_i)}_{\mu}V^{(\alpha\cap\alpha_j)}_{\nu}
\,\phi(\alpha_i)\phi(\alpha_j).
\label{eq:scalarkin}
\end{equation}
Since \eqref{eq:deriv} implies
$2V(\alpha)\hat\partial_\mu x(\alpha)
=\sum_iV^{(\alpha\cap\alpha_i)}_\mu x(\alpha_i)$, the double sum factorizes
and, for any two functions $x,y$ on the simplices,
\begin{equation}
\sum_{i,j}V^{(\alpha\cap\alpha_i)}_{\mu}V^{(\alpha\cap\alpha_j)}_{\nu}
x(\alpha_i)y(\alpha_j)
=4V(\alpha)^2\,\hat\partial_\mu x(\alpha)\hat\partial_\nu y(\alpha).
\label{eq:cellidentity}
\end{equation}
Equation~\eqref{eq:cellidentity} allows \eqref{eq:scalarkin} to be written
equivalently as
$\frac12\sum_\alpha V(\alpha)g^{\mu\nu}\hat\partial_\mu\phi
\hat\partial_\nu\phi$. The kinetic term couples simplices sharing only a
hinge, a feature of the coordinate formulation that we will revisit in
Sec.~\ref{sec:conformal}.

Higher-curvature actions follow from the replacement
\begin{equation}
\int\sqrt g\,f(R)\;\longrightarrow\;\sum_\alpha V(\alpha)f(R(\alpha)).
\end{equation}
As usual, this is
equivalent, upon introducing an auxiliary field $\Phi(\alpha)$, to
$\sum_\alpha V(\alpha)[\Phi R-W(\Phi)]$, because
\begin{equation}
f(R)=\Phi R-W(\Phi),
\qquad \Phi=f'(R),\qquad W(\Phi)=\Phi R-f(R).
\label{eq:fRaux}
\end{equation}
Indeed, varying $\int\sqrt g\,[\Phi R-W(\Phi)]$ with respect to $\Phi$ gives
$R=W'(\Phi)$, with solution $\Phi=W'^{-1}(R)$. Substituting back and using
$\Phi'(R)=f''(R)$,
\begin{equation}
W'(\Phi)=\frac{\dd W/\dd R}{\dd\Phi/\dd R}
=\frac{f''(R)R+f'(R)-f'(R)}{f''(R)}=R ,
\end{equation}
so that $\Phi R-W(\Phi)=f(R)$. Endowing $\Phi$ with a kinetic term one
obtains the Brans--Dicke action. 

\subsection{Conformal transformations and the Regge lattice}\label{sec:conformal}

Conformal transformations are central to the $f(R)$ Einstein-frame
correspondence, so we record how they act on a Regge metric.
Under the conformal transformation $g_{ab}\mapsto\tilde g_{ab}=\Omega^2g_{ab}$, with $w=\ln\Omega$, the Christoffel
symbols become
\begin{equation}
\tilde\Gamma^{a}_{bc}=\Gamma^{a}_{bc}
+\delta^{a}_{b}\partial_{c}w+\delta^{a}_{c}\partial_{b}w-g_{bc}\partial^{a}w .
\label{eq:connconf}
\end{equation}
For later use we record the conformal transformation of the curvature scalar,
\begin{equation}
R[\Omega^2g_{ab}]=\Omega^{-2}\big[R-2(d-1)\Box w-(d-1)(d-2)(\partial w)^2\big],
\label{eq:confR}
\end{equation}
where $\Box$ and $\partial$ are those of $g_{ab}$. The inverse relation is
\begin{equation}
R=\Omega^{2}\big[\tilde R+2(d-1)\tilde\Box w-(d-1)(d-2)(\tilde\partial w)^2\big].
\label{eq:RbarConf}
\end{equation}
For $d=4$ these read $\tilde R=\Omega^{-2}[R-6\Box w-6(\partial w)^2]$ and
$R=\Omega^2[\tilde R+6\tilde\Box w-6(\tilde\partial w)^2]$, the latter being
the standard relation used in the $f(R)$-Einstein-frame correspondence
\cite{Kumar2024}.

Let $g_R$ be a Regge metric, flat on the interior of each simplex, so in the affine coordinates of that simplex $\Gamma^a_{bc}=0$ and \eqref{eq:connconf} becomes
\begin{equation}
\tilde\Gamma^{a}_{bc}=\delta^{a}_{b}\partial_{c}w+\delta^{a}_{c}\partial_{b}w
-g_{bc}\partial^{a}w .
\label{eq:connflat}
\end{equation}
 For $\tilde g_{ab}=\Omega^2g_{ab}$ with $g_{ab}$ flat, flatness of
$\tilde g_{ab}$ requires the vanishing of the Ricci tensor,
\begin{equation}
\tilde R_{ab}=-(d-2)(\partial_a\partial_bw-\partial_aw\partial_bw)
-\delta_{ab}\big[\partial^2 w+(d-2)(\partial w)^2\big]=0 .
\label{eq:Riccifl}
\end{equation}
For $d\ge3$ the solutions of \eqref{eq:Riccifl} are the special conformal
(M\"obius, or inversion-type) factors, for example
$w=-2\ln|x-x_0|$, which form a measure-zero set within the space of smooth
functions. Hence a generic smooth conformal factor maps $\tilde g$ out
of the Regge class. Moreover, even the special conformal factors that do
preserve flatness do not preserve the Regge geometry, since the deficit
angles are not conformally invariant, as is shown in Appendix \ref{AnnA}.
A conformal duality therefore cannot hold as an identity between two Regge-type lattice theories.

The extended nonlocal (hinge-sharing) coupling in \eqref{eq:scalarkin} is consistent
with this conformal non-invariance. The conformal mode cannot be represented
by a local, face-sharing coupling, because a conformal transformation is
non-local on the Regge lattice. For this reason we shall not define the
discrete theory by discretizing $f(R)$ and then performing a conformal
transformation. Instead we take the Einstein-frame lattice action of
Sec.~\ref{sec:lattice} as the definition. The function $f(R)$ is used only to
fix the scalar potential.


\section{Axial torsion and the axion}\label{sec:duality}

We work with the metric-affine $f(R)$ theory with torsion of
Ref.~\cite{Kumar2024}. The connection is independent, and its curvature
scalar is $A=R(\Gamma)=g^{ab}R_{ab}(\Gamma)$. We write $R$ for $A$ when no
confusion arises, so that $f(R)$ and $f(A)$ denote the same function.
Related metric-affine formulations are discussed in
Refs.~\cite{Capozziello2007,Inagaki2022,Montesinos2020}, and the metric
$f(R)$ theory is reviewed in Ref.~\cite{Sotiriou2010}.

 The torsion and the contorsion tensors are respectively
\begin{equation}
T^{a}{}_{bc}=\Gamma^{a}_{bc}-\Gamma^{a}_{cb},
\qquad
K_{abc}=\tfrac12\big(T_{abc}-T_{bac}-T_{cab}\big).
\label{eq:Kdef}
\end{equation}
with $K_{abc}=-K_{bac}$.
Throughout this section $\Omega^2=f'(A)$ and $w=\ln\Omega$. The scalaron is
taken with the canonical normalization
\begin{equation}
\phi=\frac{\sqrt6}{\kappa}\,w,
\qquad
\Omega^2=e^{2w}=e^{2\kappa\phi/\sqrt6}.
\end{equation}

\subsection{The Einstein frame}
The Jordan-frame action for $f(R)$ gravity is
\begin{equation}S_J=\frac{1}{2\kappa^2}\int\dd^4x\sqrt{-g}\,f(R)=\frac{1}{2\kappa^2}\int\sqrt{-g}\,[\Phi R-W(\Phi)]
\end{equation}
with
$\Phi=\Omega^2=f'(A)$ and $W(\Phi)=\Phi R-f(R)$ [cf. Eq.~\eqref{eq:fRaux}]. 
Under the conformal transformation, the integration
measure transforms as $\sqrt{-g}=\Omega^{-4}\sqrt{-\tilde g}$ in $d=4$, since
$\det\tilde g_{ab}=\Omega^8\det g_{ab}$. Adding the torsion-dependent terms obtained
from the Palatini identity together with the conformal transformation of the
connection, one finds \cite{Kumar2024}
\begin{equation}
R(\Gamma)=\Omega^2\Big[\tilde{\bar R}+6\tilde\Box w-6(\tilde\partial w)^2
-2\partial_ew\,K^{a}{}_{ea}
+\Omega^{-2}K^{a}{}_{ea}K^{ed}{}_{d}
-\Omega^{-2}K_{aed}K^{eda}\Big].
\label{eq:Rfull}
\end{equation}
 The
$6\tilde\Box w$ term is a total derivative. Dropping the total derivative and
including the potential, the Einstein-frame action is
\begin{multline}
S_E=\frac{1}{2\kappa^2}\int \dd^4x\sqrt{-\tilde g}\Big[
\tilde{\bar R}-6\,\tilde g^{\mu\nu}\partial_\mu w\partial_\nu w
-2\partial_ew\,K^{a}{}_{ea}
+\Omega^{-2}K^{a}{}_{ea}K^{ed}{}_{d}\\
-\Omega^{-2}K_{aed}K^{eda}
-\frac{Af'(A)-f(A)}{f'(A)^2}\Big],
\label{eq:SEpre}
\end{multline}
with all contractions taken with $\tilde g$.

The contorsion is non-dynamical, and its equation of motion following from
\eqref{eq:SEpre} is algebraic. Its trace is
\begin{equation}
K^{a}{}_{ea}=4\,\partial_ew,
\label{eq:Ktrace}
\end{equation}
and the general solution is
\begin{equation}
K_{abc}=\frac{4}{3\Omega^2}\big(\tilde g_{ac}\partial_bw-\tilde g_{bc}
\partial_aw\big)+\partial_{[b}B_{ac]}
\equiv K^{(w)}_{abc}+K^{(B)}_{abc},
\label{eq:Ksol}
\end{equation}
where $B_{ac}=-B_{ca}$ and $B_{ac}\tilde g^{ac}=0$. This key point
of the decomposition is that the torsion is not stored in the connection
independently. The trace part is slaved to the scalaron, and the only
independent torsion that survives is axial and is carried entirely by
$B_{ac}$ through $K^{(B)}=\dd B$. It is this $B$-dependent part that becomes
the Kalb--Ramond two-form below. The trace in
\eqref{eq:Ktrace} is contracted with $g$, and the coefficient in
\eqref{eq:Ksol} is written with $\tilde g$ so that
$(4/(3\Omega^2))\tilde g_{ac}=(4/3)g_{ac}$. Two structural properties will be our main tools. The
first is that \begin{equation}K^{(w)}_{[abc]}=0,\end{equation} so that the vector part has no totally
antisymmetric component. The second is that
$K^{(B)}_{abc}=\partial_{[b}B_{ac]}$ is totally antisymmetric and traceless, so that the trace \eqref{eq:Ktrace} is carried entirely by $K^{(w)}$.

Writing $K=K^{(w)}+K^{(B)}$, the $B$-dependent part of \eqref{eq:SEpre}
becomes
\begin{equation}
-\Omega^{-2}K_{aed}K^{eda}
=-\Omega^{-2}\Big(K^{(w)}_{aed}K^{(w)eda}
+2K^{(w)}_{aed}K^{(B)eda}
+K^{(B)}_{aed}K^{(B)eda}\Big).
\label{eq:expand}
\end{equation}
The mixed term vanishes,
\begin{equation}
K^{(w)}_{aed}K^{(B)eda}=0 ,
\label{eq:crosszero}
\end{equation}
because contracting a tensor that is antisymmetric in two indices with a
totally antisymmetric tensor leaves only the totally antisymmetric part of
the former. Since $K^{(w)}_{[abc]}=0$ and $K^{(B)}$ is totally antisymmetric,
the contraction must vanish. Hence the torsion sector is a \emph{single} quadratic
form.

\subsection{Torsional action}
Let us introduce the canonically normalized two-form and the dressed field strength
\begin{equation}
Z_{ab}=\frac{\Omega^2}{\kappa}B_{ab},
\qquad
\mathcal H_{abc}:=\partial_{[a}Z_{bc]}-2\,\partial_{[a}w\,Z_{bc]},
\label{eq:Hdressed}
\end{equation}
the latter being the totally antisymmetric component of
$\dd Z-2\,\dd w\wedge Z$. From $B=(\kappa/\Omega^2)Z$ one has
$\partial_{[a}B_{bc]}=\kappa\Omega^{-2}\mathcal H_{abc}$. The $B$-dependent
part of the action is therefore
\begin{equation}
S_{\rm tor}=\frac{1}{2\kappa^2}\big(-\Omega^{4}\big)\,\kappa^2\Omega^{-4}
\int\dd^4x\sqrt{-\tilde g}\,\big|\mathcal H\big|^2
=-\frac12\int\dd^4x\sqrt{-\tilde g}\,\big|\mathcal H\big|^2,
\label{eq:Stor}
\end{equation}
with $|\mathcal H|^2=\mathcal H_{abc}\mathcal H^{abc}$.
The factor $\Omega^4$ in \eqref{eq:Stor} is a bookkeeping convention that
assigns the conformal factor of the measure to the kinetic term. This
convention differs from Ref.~\cite{Kumar2024}, but the physical action and
the equations of motion are unchanged.
The field-dependent redefinition
\begin{equation}
Z_{ab}=\Omega^2Z'_{ab}
\label{eq:Zprime}
\end{equation}
gives $\mathcal H_{abc}=\Omega^2\partial_{[a}Z'_{bc]}$ and $B=\kappa Z'$, so
\begin{equation}
S_{\rm tor}=-\frac12\int\dd^4x\sqrt{-\tilde g}\,
\Omega^4\,\big|\dd Z'\big|^2,
\qquad \Omega^4=e^{4w},
\label{eq:dilatonKR}
\end{equation}
which is a standard dilaton--Kalb--Ramond action \cite{Kalb1974,Scherk1974}. Here $Z'$ carries the reducible
two-form gauge symmetry $\delta Z'_{ab}=\partial_{[a}\lambda_{b]}$ with
$\lambda_a\to\lambda_a+\partial_a\varepsilon$.

Promoting $H'=\dd Z'$ to an independent three-form and enforcing $\dd H'=0$
by a Lagrange multiplier $a$,
\begin{equation}
S=\int\Big[-\frac12\,\Omega^4\,H'\wedge\star H'+a\,\dd H'\Big],
\label{eq:dualint}
\end{equation}
where the form norm $H'\wedge\star H'$ differs from the component norm of
\eqref{eq:dilatonKR} by a positive numerical factor absorbed in the
conventional normalization. Varying with respect
to $H'$ gives $\Omega^4\star H'=-\dd a$, that is,
\begin{equation}
H'=\sigma'\,\Omega^{-4}\star\dd a ,
\label{eq:Hdual}
\end{equation}
and substituting back,
\begin{equation}
S_a=\frac{\sigma}{2}\int\dd^4x\sqrt{-\tilde g}\,
\Omega^{-4}\,(\partial a)^2,
\qquad \Omega^{-4}=e^{-4w},
\label{eq:Saxion}
\end{equation}
with $\sigma=\pm1$. In the Euclidean convention of the lattice
$(\partial a)^2>0$, so $\sigma=+1$ gives a positive (healthy) kinetic term.\footnote{In the Lorentzian signature of \eqref{eq:Eframe}, the combination
$(\partial a)^2=-(\partial_ta)^2+(\partial_ia)^2$ is negative for a
time-dependent axion, so the healthy choice is $\sigma=-1$. This gives
$S_a=-\tfrac12\int\sqrt{-\tilde g}\,\Omega^{-4}(\partial a)^2$ and matches
the axion term of \eqref{eq:Eframe}. The Wick rotation of
Sec.~\ref{sec:frw} connects the two conventions.} This establishes the
equivalence that the torsion sector is exactly a single massless scalar, the
axion, with the field-dependent kinetic term \eqref{eq:Saxion}.

With $A^d=\tfrac{1}{3!}\epsilon^{dabc}T_{[abc]}$ and
$H'_{abc}\propto\epsilon_{abcd}\partial^da$,
\begin{equation}
A^d\propto\epsilon^{dabc}H'_{abc}
=\epsilon^{dabc}\epsilon_{abce}\partial^ea
=-3!\,\delta^d_e\,\partial^ea
=-6\,\partial^da ,
\end{equation}
so that, up to normalization, $A_\mu=\partial_\mu a$,
\begin{equation}
A_\mu=\partial_\mu a .
\label{eq:Agrad}
\end{equation}
Torsion-free configurations are therefore exactly those with a constant
axion. The trace part of the contorsion is absorbed into the scalaron and the
tensor part is removed algebraically.

In four dimensions the axial torsion, equivalently the two-form field
strength, carries the degrees of freedom of a massless two-form, dual to a
single scalar. Adding the scalaron, the total bosonic content is a massless
graviton with two polarizations, a scalaron, and an axion, so that
$2+1+1=4$ degrees of freedom. 

\subsection{Comparison with other simplicial torsion formulations}
\label{sec:comparison}

The on-shell contorsion \eqref{eq:Ksol} contains only a vector and an
axial part, $K=K^{(w)}+K^{(B)}$, the tensor part being absent. Since
$K_{[abc]}=\tfrac12T_{[abc]}=K^{(B)}_{abc}$, the axial torsion is
\begin{equation}
\tfrac12\,T_{[abc]}=\partial_{[b}B_{ac]}
=\kappa\,\partial_{[b}Z'_{ac]}
=\kappa\,\mathcal H'_{abc},
\label{eq:axialproj}
\end{equation}
with $B=\kappa Z'$, so that $\mathcal H'=\dd Z'$ equals
$T_{[abc]}/(2\kappa)$, that is, $1/(2\kappa)$ times the axial torsion.

Torsion-free configurations are equivalent to $\dd B=0$ with
$\partial_\mu w=0$, or to $\dd a=0$ and $\partial_\mu\phi=0$ in the Einstein
frame, that is, $f(R)=R+\Lambda$. The conditions
$\dd e^a-\omega^{ab}\wedge e_b=0$ \cite{Xue2009} and the edge-vector
matching \cite{Yan2026} are both statements that the torsion two-form
vanishes. The axial projection of any torsion two-form is the same
three-form, so the torsion here is the axial projection of the first-order
and holonomy formulations. Table~\ref{tab:comparison} summarizes the
comparison. 
\begin{table}[t]
\centering
\footnotesize
\renewcommand{\arraystretch}{1.2}
\setlength{\tabcolsep}{4pt}
\caption{Comparison of simplicial torsion formulations.}
\label{tab:comparison}
\begin{tabular}{@{}>{\raggedright\arraybackslash}p{2.4cm}>{\raggedright\arraybackslash}p{3.7cm}>{\raggedright\arraybackslash}p{3.9cm}>{\raggedright\arraybackslash}p{3.9cm}@{}}
\hline
& Geometric defects \cite{Drummond1986,SchmidtKohler2001,PereiraVargas2002}
& Holonomy and first order \cite{Xue2009,Yan2026}
& This work \\
\hline
Carrier & Burgers vector on hinges and body torsion in simplices
& two-form $T^I=\dd e^I+\omega^I{}_J\wedge e^J$ on triangles
& three-form $H'=\dd Z'$ on three-cells dual to edges \\
Physical meaning & closure failure (dislocation) &
mismatch of vielbeins across an interface &
field strength of a two-form gauge field \\
Torsion-free & vanishing Burgers vector &
edge-vector matching \cite{Yan2026} and $\dd e^a-\omega^{ab}\wedge e_b=0$
\cite{Xue2009} & $H'=0$, that is, $a=$ const \\
Sector & full torsion & full torsion & axial only \\
Matter & Burgers vector couples algebraically to spin
\cite{SchmidtKohler2001} & holonomy coupling &
coupling mediated by the scalaron \\
Gauge & local rotations & local Lorentz holonomy &
two-form symmetry $Z'\to Z'+\dd\lambda$ \\
\hline
\end{tabular}

\end{table}

\section{Regge discretization}\label{sec:lattice}

Throughout this section the continuum theory is the Einstein-frame theory with the total action
\begin{equation}
S=\int \dd^4x\sqrt{-\tilde g}\Big[\frac{\tilde R}{2\kappa^2}
-\frac12(\partial\phi)^2-W(\phi)
-\frac12\Omega^{-4}(\phi)\,(\partial a)^2\Big],
\qquad \Omega^2=f'(A),
\label{eq:Eframe}
\end{equation}
where $\phi=\sqrt6\,w/\kappa$ and
$\Omega^{-4}(\phi)=e^{-4w}=e^{-4\kappa\phi/\sqrt6}$.

\subsection{The lattice action}

Let $\phi(\alpha)$ and $a(\alpha)$ be scalar fields on the dual sites, the
simplices. Applying \eqref{eq:dictionary} term by term to \eqref{eq:Eframe},
and using the Euclidean convention of Sec.~\ref{sec:regge}, we obtain
\begin{equation}
\begin{aligned}
S_{\rm latt}=\;&\frac{1}{2\kappa^2}\sum_{h}V(h)\sin\theta_h
+\sum_{\alpha}V(\alpha)\,W(\phi(\alpha))\\[2pt]
&+\frac{1}{8}\sum_{\alpha}\frac{g^{\mu\nu}(\alpha)}{V(\alpha)}
\sum_{i,j=1}^{d+1}V^{(\alpha\cap\alpha_i)}_{\mu}(\alpha)\,
V^{(\alpha\cap\alpha_j)}_{\nu}(\alpha)\,
\Big[\phi(\alpha_i)\phi(\alpha_j)
+\Omega^{-4}(\phi(\alpha))\,a(\alpha_i)a(\alpha_j)\Big],
\end{aligned}
\label{eq:Slatt}
\end{equation}
with
\begin{equation}
W(\phi)=\frac{Af'(A)-f(A)}{2\kappa^2 f'(A)^2}\Big|_{A=A(\phi)},
\qquad \Omega^{-4}(\phi)=e^{-4w}.
\end{equation}
This  Euclidean lattice action \eqref{eq:Slatt} comes from the following reasoning.
The Einstein-Hilbert term becomes $\frac{1}{2\kappa^2}\sum_hV(h)\sin\theta_h$
by \eqref{eq:EHdisc}. The scalar kinetic term gives
\begin{align}
\frac12\sum_\alpha V(\alpha)\,g^{\mu\nu}\hat\partial_\mu\phi\hat\partial_\nu\phi
&=\frac18\sum_\alpha
\frac{g^{\mu\nu}(\alpha)}{V(\alpha)}
\sum_{i,j}V^{(\alpha\cap\alpha_i)}_{\mu}V^{(\alpha\cap\alpha_j)}_{\nu}
\,\phi(\alpha_i)\phi(\alpha_j),
\label{eq:kinderiv}
\end{align}
which fixes the factor $1/8$. The potential gives
$\sum_\alpha V(\alpha)W(\phi(\alpha))$, and the axion kinetic term differs
from the scalar kinetic term only by the field-dependent prefactor
$\Omega^{-4}(\phi)$ evaluated at the simplex.

Setting $H'=0$, equivalently $a=$ const, makes the axion kinetic term vanish, so that 
\begin{equation}
S_{\rm latt}\big|_{H'=0}
=\frac{1}{2\kappa^2}\sum_hV(h)\sin\theta_h
+\sum_\alpha V(\alpha)\Big[\tfrac12\,g^{\mu\nu}\hat\partial_\mu\phi
\hat\partial_\nu\phi+W(\phi)\Big],
\label{eq:torsionfreelimit}
\end{equation}
which is Regge gravity coupled to the scalaron. For a smooth configuration
$S_{\rm latt}=S+\mathcal O(l^2)$ by the second-order accuracy
\eqref{eq:derivcont} and \eqref{eq:EHdisc}. In the $H'=0$ limit only the
\emph{axial} torsion is switched off. The trace part $K^{(w)}$ is slaved to
the scalaron and propagates no degree of freedom.

\subsection{Equations of motion}\label{ss:eom}

The lattice action \eqref{eq:Slatt} depends on three independent sets of
variables, namely the vertex scalars $\phi(\alpha)$ and $a(\alpha)$ and the
edge lengths $l_{ij}$. Varying term by term we obtain three equations, which
parallel the two variations (edge vectors and holonomies) of the simplicial
Einstein-Cartan formulation of Ref.~\cite{Yan2026}.

\paragraph{(i) Axion.} Varying with respect to $a(\gamma)$ gives the discrete
current conservation
\begin{equation}
\sum_{\beta\succ\gamma}\Omega^{-4}(\phi(\beta))\,
V^{(\beta\cap\gamma)}_{\mu}(\beta)\,g^{\mu\nu}(\beta)\,
\hat\partial_\nu a(\beta)=0 ,
\label{eq:eom_a}
\end{equation}
the discrete analogue of $\nabla_a j^a=0$ with
$j^a=\Omega^{-4}\partial^a a$.

\paragraph{(ii) Scalaron.} Varying with respect to $\phi(\gamma)$ gives the
discrete Klein--Gordon equation
\begin{equation}
\hat\Box\phi(\gamma)-W'(\phi(\gamma))
+\frac{1}{2}\frac{\dd\Omega^{-4}}{\dd\phi}(\phi(\gamma))\,
g^{\mu\nu}(\gamma)\hat\partial_\mu a(\gamma)\hat\partial_\nu a(\gamma)=0 ,
\label{eq:eom_phi}
\end{equation}
with $\frac{\dd\Omega^{-4}}{\dd\phi}=-\frac{4\kappa}{\sqrt6}\Omega^{-4}$.

\paragraph{(iii) Edge equation.} This is the counterpart of the variation of
the edge vectors in Ref.~\cite{Yan2026}. The variation of the Regge term is
\begin{equation}
\delta S_R=\frac{1}{2\kappa^2}\sum_h\Big[\sin\theta_h\,\delta V(h)
+V(h)\cos\theta_h\,\delta\theta_h\Big],
\label{eq:dSR}
\end{equation}
and the Sch\"afli identity,
\begin{equation}
\sum_{h\subset\sigma}V(h)\,\delta\theta^{(\sigma)}_h=0 ,
\label{eq:schlafli}
\end{equation}
which for $d=2$ is the constancy of the sum of a triangle's interior angles,
for $d=3$ reads $\sum_e\ell_e\,\delta\theta_e=0$, and for $d=4$ reads
$\sum_t A_t\,\delta\theta_t=0$. Summing \eqref{eq:schlafli} over all
$4$-simplices and using
$\theta_h=2\pi-\sum_{\sigma\supset h}\theta^{(\sigma)}_h$ gives
\begin{equation}
\sum_h V(h)\,\delta\theta_h
=-\sum_\sigma\sum_{h\subset\sigma}V(h)\,\delta\theta^{(\sigma)}_h=0
\label{eq:schlafli2}
\end{equation}
for arbitrary edge variations. Writing $\cos\theta_h=1-2\sin^2(\theta_h/2)$,
the $\theta$ variation in \eqref{eq:dSR} is then of order
$\theta_h^2=O(l^4)$ relative to the first term, the absolute correction
being $O(l^7)$ per hinge, so that to the working accuracy $O(l^2)$
\begin{equation}
\frac{\delta S_R}{\delta l_e}
=\frac{1}{2\kappa^2}\sum_{h\supset e}\sin\theta_h\,
\frac{\partial V(h)}{\partial l_e}\Big[1+O(l^4)\Big],
\end{equation}
and $\delta S_{\rm latt}/\delta l_e=0$ reads
\begin{equation}
\sum_{h\supset e}\sin\theta_h\,\frac{\partial V(h)}{\partial l_e}
+2\kappa^2\,\frac{\partial S_{\rm m}}{\partial l_e}=0 ,
\label{eq:edgeEOM}
\end{equation}
where $S_{\rm m}=S_\phi+S_a$. The matter variation receives contributions
from the volume factor and from the kinetic terms,
\begin{equation}
\frac{\partial S_{\rm m}}{\partial l_e}
=\sum_\alpha\Big[\frac{\partial V(\alpha)}{\partial l_e}\,
\mathcal L_{\rm m}(\alpha)
+V(\alpha)\,\frac{\partial\mathcal L_{\rm kin}(\alpha)}{\partial l_e}\Big],
\label{eq:dSm}
\end{equation}
with $\mathcal L_{\rm m}=\mathcal L_{\rm kin}+W$ and
$\mathcal L_{\rm kin}=\tfrac12 g^{\mu\nu}\hat\partial_\mu\phi
\hat\partial_\nu\phi+\tfrac12\Omega^{-4}g^{\mu\nu}
\hat\partial_\mu a\hat\partial_\nu a$. The derivative in the second term
acts both on $g^{\mu\nu}(\alpha)$ and on the lattice derivative
$\hat\partial_\mu$. Defining
$G_e:=\sum_{h\supset e}\sin\theta_h\,\partial V(h)/\partial l_e$ and
$T_e:=-2\,\partial S_{\rm m}/\partial l_e$, the edge equation is
$G_e=\kappa^2 T_e$. It involves \emph{all} hinges sharing the edge $e$.

The Regge part of \eqref{eq:edgeEOM} is the standard Regge equation and, in
the continuum limit $l\to0$, becomes the Einstein tensor. Thus
\eqref{eq:edgeEOM} reproduces the Einstein equation of \eqref{eq:Eframe},
\begin{equation}
G_{ab}=\kappa^2\big(T^{(\phi)}_{ab}+T^{(a)}_{ab}\big),
\end{equation}
with
\begin{align}
T^{(\phi)}_{ab}&=\partial_a\phi\partial_b\phi
-g_{ab}\Big[\tfrac12(\partial\phi)^2+W(\phi)\Big],\\
T^{(a)}_{ab}&=\Omega^{-4}\Big[\partial_a a\,\partial_b a
-\tfrac12 g_{ab}(\partial a)^2\Big].
\end{align}
The divergence of \eqref{eq:edgeEOM} is equivalent to (i) and (ii), the
discrete analogue of $\nabla_a T^{ab}=0$. 

We remark that, in this construction, the geometry is still carried entirely by the edge lengths,
while the scalaron $\phi$ and the axion $a$ are fields on the dual vertices.
This division is the lattice counterpart of the field redefinition that maps
the Jordan frame to the Einstein frame. A nonzero axion gradient is the
axial torsion, but it does not bend the simplices or require a local frame.
Its backreaction on the geometry is through the stress-energy $T_e$ in the
edge equation, where it acts as a stiff fluid. The scalaron carries the
conformal mode. It does not rescale each simplex independently, since a
generic conformal factor would take the metric out of the Regge class, but
enters the edge equation and the homogeneous equations of
Sec.~\ref{sec:frw} through its potential and kinetic term. In this sense the
effects of torsion and of the conformal factor are put entirely into the
fields, and the geometry responds only collectively through the edge
equations. This is the opposite of the geometric approach in the literature, in which torsion
is stored in dislocations or in a local frame on each simplex, and it is the
reason the present construction needs no geometric torsion.

\section{Homogeneous and isotropic cosmology}\label{sec:frw}

We now apply the discrete equations of motion to a homogeneous and isotropic
configuration. The detailed derivation of this section is given in
App.~\ref{app:frw}. Regge cosmology was initiated by Collins and Williams
\cite{Collins1973}, whose 5-, 16-, and 600-cell models of the closed
Friedmann universe remain the standard regular triangulations. Refined and
Lorentzian versions are given in, for example, \cite{Liu2016,Tsuda2021,Jercher2024},
and simplicial minisuperspace models go back to Hartle \cite{Hartle1985}.

\subsection{The homogeneous lattice ansatz}

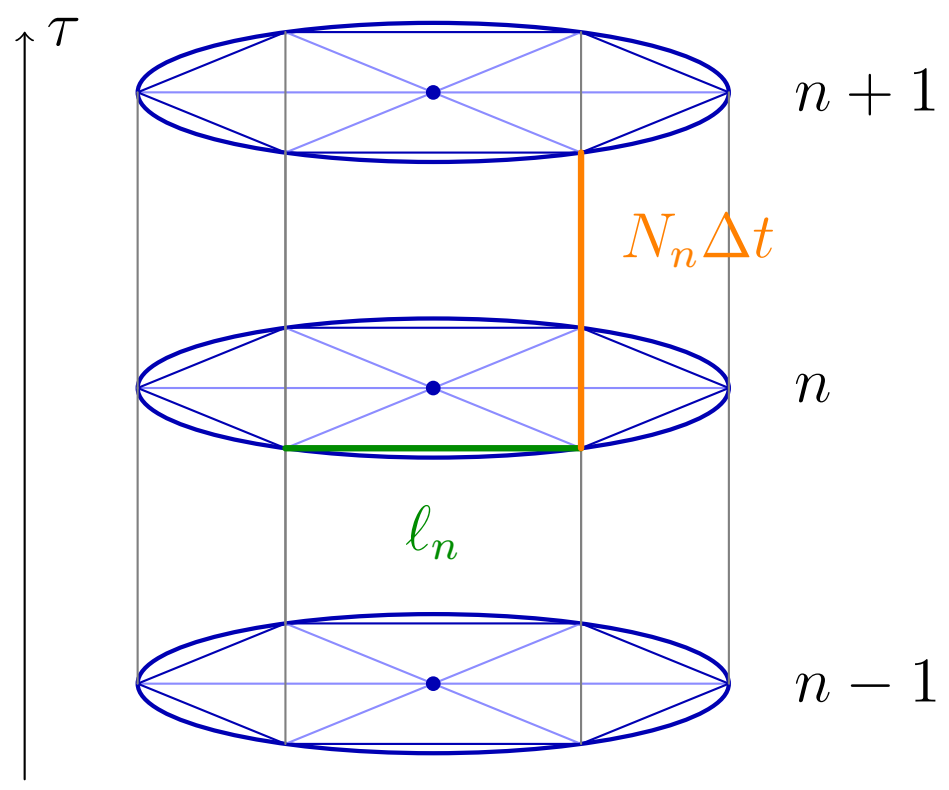
\begin{figure}[t]
\centering
\begin{tikzpicture}[scale=1.15,line join=round,line cap=round]
  \def\rx{1.7}
  \def\ry{0.4}
  \newcommand{\slice}[2]{%
    \draw[blue!70!black,thick] (0,#1) ellipse ({\rx} and {\ry});
    \foreach \a in {0,60,...,300} {
      \coordinate (#2\a) at ({\rx*cos(\a)}, {#1+\ry*sin(\a)});
      \draw[blue!45] (0,#1) -- (#2\a);
    }
    \draw[blue!70!black] (#20)--(#260)--(#2120)--(#2180)--(#2240)--(#2300)--cycle;
    \fill[blue!70!black] (0,#1) circle (1.2pt);
  }
  \slice{0}{A}
  \slice{1.7}{B}
  \slice{3.4}{C}
  \foreach \a in {0,60,...,300} {
    \draw[gray] (A\a)--(B\a);
    \draw[gray] (B\a)--(C\a);
  }
  \draw[very thick,green!55!black] (B240)--(B300);
  \draw[very thick,orange] (B300)--(C300);
  \node[below,green!55!black] at (0,1.7-0.55) {$\ell_n$};
  \node[right,orange] at (0.95,2.55) {$N_n\Delta t$};
  \node[right] at (1.95,0) {$n-1$};
  \node[right] at (1.95,1.7) {$n$};
  \node[right] at (1.95,3.4) {$n+1$};
  \draw[->] (-2.35,-0.55)--(-2.35,3.75) node[right] {$\tau$};
\end{tikzpicture}
\caption{Schematic layered triangulation of the cosmological lattice. Each
time slice carries a regular triangulation of the spatial section, with all
spatial edges equal to $\ell_n$. Consecutive slices are joined by timelike
edges of length $N_n\Delta t$, and $\tau$ is proper time. The drawing is a
two-dimensional projection; the actual spatial sections are the triangulations of $\mathbb S^3$.}
\label{fig:lattice}
\end{figure}

We take a layered triangulation of a closed or flat spatial slice
(Fig.~\ref{fig:lattice}). Within
each time slice $n$ all spatial edges are equal to $\ell_n$, the scale factor
in lattice units, and consecutive slices are joined by timelike edges of
length $N_n\,\Delta t$, where $N_n$ is the lapse. The scalars are constant on
each slice. Absorbing the overall spatial-cell count into a common
normalization, the lattice action \eqref{eq:Slatt} reduces to the
minisuperspace form, Lorentzian after the standard Wick rotation of the
Euclidean lattice,
\begin{equation}
\begin{aligned}
S_{\rm ms}=\sum_n\Big[&
-\frac{3\ell_n(\Delta\ell_n)^2}{\kappa^2N_n\Delta t}
+\frac{3N_nk\ell_n\Delta t}{\kappa^2}
+\frac{\tfrac12\ell_n^3(\Delta\phi_n)^2}{N_n\Delta t}\\
&-N_n\Delta t\,\ell_n^3 W(\phi_n)
+\frac{\tfrac12\Omega^{-4}(\phi_n)\ell_n^3(\Delta a_n)^2}{N_n\Delta t}
\Big],
\end{aligned}
\label{eq:Sms}
\end{equation}
with $\Delta q_n:=q_{n+1}-q_n$ and $k$ the unit curvature of the spatial slices. The first two terms are the homogeneous reduction of the
Regge term, namely its extrinsic- and intrinsic-curvature parts, and the
remaining terms are the homogeneous reduction of $S_\phi$ and $S_a$. It is
convenient to introduce
\begin{equation}
H_n:=\frac{\Delta\ell_n}{\ell_nN_n\Delta t},
\qquad
\dot\phi_n:=\frac{\Delta\phi_n}{N_n\Delta t},
\qquad
\dot a_n:=\frac{\Delta a_n}{N_n\Delta t}.
\end{equation}

Varying \eqref{eq:Sms} with respect to the lapse $N_n$ gives the discrete
Hamiltonian constraint, that is, the discrete Friedmann equation,
\begin{equation}
3H_n^2=\kappa^2\Big[\tfrac12\dot\phi_n^2+W(\phi_n)
+\tfrac12\Omega^{-4}(\phi_n)\,\dot a_n^2\Big]
-\frac{3k}{\ell_n^2} .
\label{eq:discFried}
\end{equation}
Varying with respect to $\ell_n$ gives the discrete acceleration
(Raychaudhuri) equation,
\begin{multline}
\frac{(\Delta\ell_n)^2-2\ell_n\Delta\ell_n}{N_n}
+\frac{2\ell_{n-1}\Delta\ell_{n-1}}{N_{n-1}}
=\kappa^2\Big[\tfrac12\ell_n^2\frac{(\Delta\phi_n)^2}{N_n}\\
+\tfrac12\Omega^{-4}\ell_n^2\frac{(\Delta a_n)^2}{N_n}
-N_n\Delta t^2\ell_n^2W(\phi_n)\Big]+kN_n\Delta t^2.
\label{eq:discRay}
\end{multline}
Varying with respect to $\phi_n$ and $a_n$ gives the discrete scalaron and
axion equations,
\begin{equation}
\frac{\ell_n^3\dot\phi_n-\ell_{n-1}^3\dot\phi_{n-1}}{\Delta t}
=N_n\ell_n^3\Big[-W'(\phi_n)
+\tfrac12\frac{\dd\Omega^{-4}}{\dd\phi}\dot a_n^2\Big]
\label{eq:discKG}
\end{equation}
and
\begin{equation}
\ell_n^3\Omega^{-4}(\phi_n)\,\dot a_n
=\ell_{n-1}^3\Omega^{-4}(\phi_{n-1})\,\dot a_{n-1} .
\label{eq:discaxion}
\end{equation}
Equation~\eqref{eq:discaxion} is the discrete form of
$\partial_\tau(\ell^3\Omega^{-4}\dot a)=0$, that is,
$\Omega^{-4}\dot a\propto\ell^{-3}$. It defines the conserved axion charge
\begin{equation}
Q:=\ell^3\Omega^{-4}\dot a ,
\label{eq:Q}
\end{equation}
which is used in the numerical examples below.

\subsection{The regular triangulation of $\mathbb S^3$}

To make the coefficients in \eqref{eq:Sms} explicit we take the spatial slice
to be a regular (equilateral) triangulation of $\mathbb S^3$ with all edges
$\ell=1$ at unit scale, $N_0$ vertices, $N_1$ edges, $N_2$ triangles, and
$N_3$ tetrahedra, each edge being shared by $n_3$ tetrahedra.  The regular tetrahedral
triangulations of $\mathbb S^3$ are the 5-, 16-, and 600-cell, the three
regular 4-polytopes with tetrahedral cells \cite{Coxeter1973,Collins1973}, and the
600-cell is the finest of them. The so-called f-vectors $(N_0,N_1,N_2,N_3)$ are
$(5,10,10,5)$, $(8,24,32,16)$, and $(120,720,1200,600)$ for the 5-, 16-, and
600-cell, respectively. Each satisfies $N_0-N_1+N_2-N_3=0$, and the
coefficients used below are $N_1=6N_3/n_3$ and $V_0=N_3/(6\sqrt2)$.

An alternative to the regular triangulations is the prism (staircase)
triangulation of $\Delta^3\times I$ \cite{DeLoera2010}, in
which each spatial tetrahedron is
extended to four $4$-simplices. It introduces cross edges $ij'$ whose lengths
are independent variables; for the straight embedding one has
$c^2=\ell^2+T^2$. The mixed-hinge sum is well defined [cf. Appendix~\ref{app:prism}]. The staircase, however, depends on an ordering of the
vertices and is not vertex-transitive.

For equilateral
tetrahedra the dihedral angle is $\theta_\Delta=\arccos(1/3)$, so the spatial
deficit angle at every edge is
\begin{equation}
\theta_{sp}=2\pi-n_3\arccos\tfrac13 ,
\qquad
V_0=\frac{N_3}{6\sqrt2},
\qquad
N_1=\frac{6N_3}{n_3},
\label{eq:thetasp}
\end{equation}
where $V_0$ is the $\ell=1$ volume of the slice. The induced curvature is
exact for the equilateral slice,
$\int\sqrt h\,{}^{(3)}R=\sum_e\ell_e\theta^{sp}_e=N_1\ell\theta_{sp}$, so
that
\begin{equation}
\begin{aligned}
\frac{1}{2\kappa^2}\int N\sqrt h\,{}^{(3)}R
&\;\longrightarrow\;
\frac{N_1\theta_{sp}}{2\kappa^2}\,N_n\,\ell_n\,\Delta t
\equiv\frac{3\mathcal K}{2\kappa^2}\,N_n\ell_n\Delta t ,\\
\mathcal K&:=\frac{N_1\theta_{sp}}{6V_0}
=\frac{6\sqrt2\,(2\pi-n_3\arccos\frac13)}{n_3},
\end{aligned}
\label{eq:Knum}
\end{equation}
while the extrinsic-curvature part becomes $-3V_0\ell\dot\ell^2/N$, which
reproduces the kinetic term of \eqref{eq:Sms} with the volume factor $V_0$
made explicit. For the regular triangulation the unit-curvature term is
replaced by $3k\mapsto N_1\theta_{sp}/(2V_0)=3\mathcal K$
[Eq.~\eqref{eq:Knum}]. Hence \eqref{eq:Sms} becomes
\begin{equation}
\begin{aligned}
S_{\rm ms}=V_0\sum_n\Big[
&-\frac{3}{\kappa^2}\frac{\ell_n(\Delta\ell_n)^2}{N_n\Delta t}
+\frac{N_1\theta_{sp}}{2V_0\kappa^2}N_n\ell_n\Delta t\\
&+\frac{\ell_n^3(\Delta\phi_n)^2}{2N_n\Delta t}
-N_n\Delta t\,\ell_n^3W(\phi_n)
+\frac{\Omega^{-4}\ell_n^3(\Delta a_n)^2}{2N_n\Delta t}\Big],
\end{aligned}
\label{eq:Smsreg}
\end{equation}
and the discrete Friedmann equation reads
\begin{equation}
3H_n^2=\kappa^2\Big[\tfrac12\dot\phi_n^2+W(\phi_n)
+\tfrac12\Omega^{-4}(\phi_n)\dot a_n^2\Big]
-\frac{3\mathcal K}{\ell_n^2} .
\label{eq:discFriedreg}
\end{equation}
  Since the discrete curvature
term $3\mathcal K/\ell_n^2$ enters \eqref{eq:discFriedreg} with the opposite
sign to the energy density, a closed lattice universe has a bounce at
\begin{equation}
\ell_{\rm bounce}=\sqrt{\frac{3\mathcal K}{\kappa^2\rho}}.
\label{eq:bounce}
\end{equation}

For a smooth configuration and a {\it flat} spatial slice,
$\Delta q_n=\dot q\,N_n\Delta t+O(\Delta t^2)$, Eqs.~\eqref{eq:discFried}--\eqref{eq:discaxion}
reduce to the Friedmann--Raychaudhuri system of \eqref{eq:Eframe}:
\begin{align}
3H^2&=\kappa^2\Big[\tfrac12\dot\phi^2+W(\phi)
+\tfrac12\Omega^{-4}\dot a^2\Big],\label{eq:contFried}\\
2\dot H+3H^2&=-\kappa^2\Big[\tfrac12\dot\phi^2-W(\phi)
+\tfrac12\Omega^{-4}\dot a^2\Big],\\
\ddot\phi+3H\dot\phi+W'(\phi)
&=\tfrac12\frac{\dd\Omega^{-4}}{\dd\phi}\dot a^2
=-\frac{2\kappa}{\sqrt6}\Omega^{-4}\dot a^2,\\
\partial_\tau\big(\Omega^{-4}\dot a\big)+3H\Omega^{-4}\dot a&=0 ,
\end{align}
with derivatives with respect to proper time. The axion behaves as a stiff
($w=1$) fluid with $\Omega^{-4}\dot a\propto\ell^{-3}$ and sources the
scalaron through the last term of the $\phi$ equation. For the Starobinsky
model $f(R)=R+R^2/(6M^2)$ the scalaron potential is
$W(\phi)=(3M^2/4\kappa^2)(1-\Omega^{-2})^2$.

\begin{figure}[t]
  \centering
  \includegraphics[width=\textwidth]{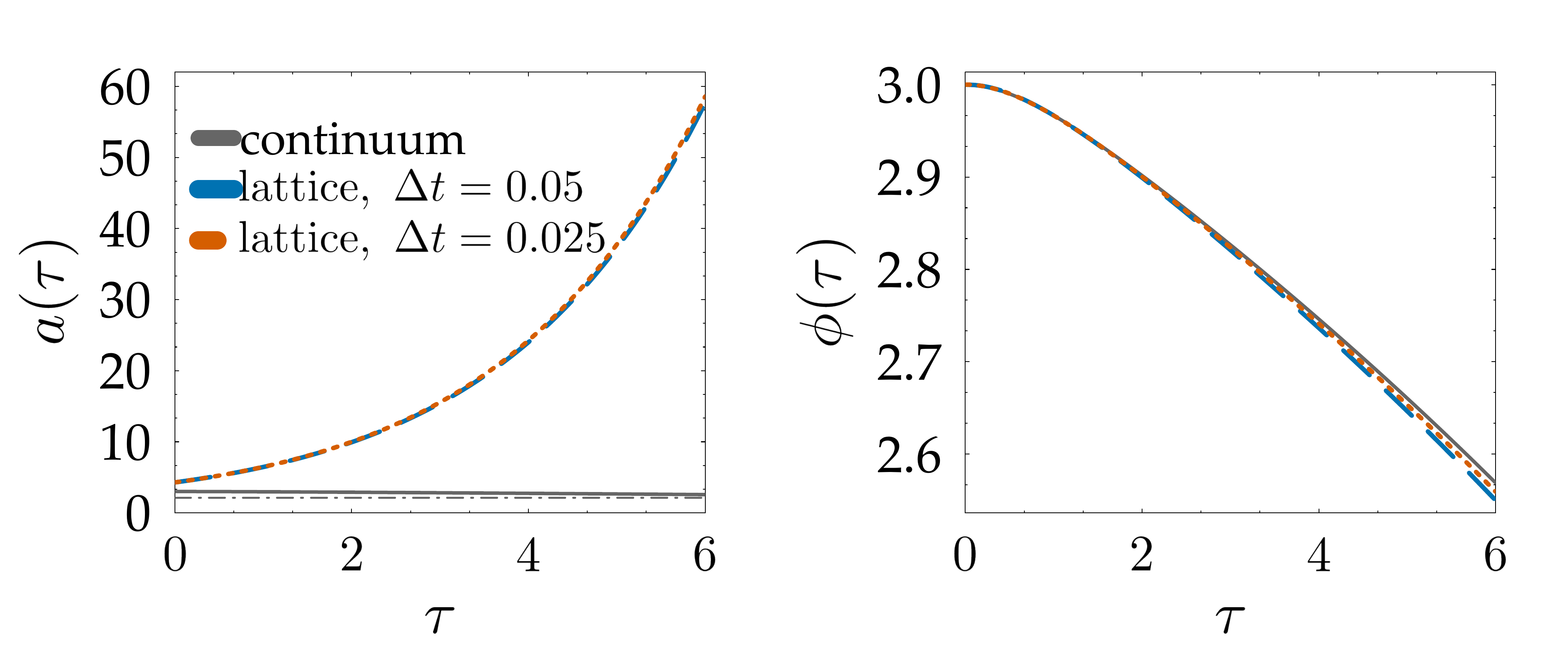}
  \caption{Homogeneous cosmology on the 600-cell lattice for the Starobinsky
  model with a nonzero axion charge $Q=0.02$, $\kappa=1$, $M=1$,
  $\phi(0)=3$, $\dot\phi(0)=0$, and $\ell(0)=2\ell_{\rm bounce}$. Left, the
  physical scale factor $a(\tau)=\ell(\tau)/\sqrt{\mathcal K}$. Right, the
  scalaron $\phi(\tau)$. The gray curves are the continuum closed solution,
  the dashed and dotted curves are the lattice solutions at $\Delta t=0.05$
  and $0.025$, and the horizontal line in the left panel marks the bounce
  radius.}
  \label{fig:evolution}
\end{figure}

For the numerical integration we work on the 600-cell triangulation, so that
$\mathcal K=0.2179$ and the curvature term in \eqref{eq:discFriedreg} is
resolved, and we include a nonzero axion charge $Q=0.02$. We start at
$\ell(0)=2\ell_{\rm bounce}$ with $\phi(0)=3$ and $\dot\phi(0)=0$. We solve
the discrete Friedmann equation \eqref{eq:discFriedreg} for the scale
factor, the scalaron equation \eqref{eq:discKG} for $\phi$, and the axion
current \eqref{eq:discaxion}. The discrete Raychaudhuri equation
\eqref{eq:discRay} is not imposed independently. We monitor its relative
residual, which decreases with $\Delta t$ in the same way as the trajectory
error. In the continuum the Raychaudhuri equation follows from the other
three. At finite $\Delta t$ the discrete equations are consistent only up to
the discretization error, so \eqref{eq:discRay} is used as a constraint
rather than an independent equation. Its residual in
Table~\ref{tab:convergence} is of the same order as the trajectory error, so
the numerical solution is an $\mathcal O(\Delta t)$ solution of the full
discrete system. Figure~\ref{fig:evolution} shows the physical scale factor
$a(\tau)=\ell(\tau)/\sqrt{\mathcal K}$ and the scalaron for the lattice and
for the continuum closed solution. The lattice curves converge to the
continuum as $\Delta t\to0$, and the bounce radius
$a_{\rm bounce}=\ell_{\rm bounce}/\sqrt{\mathcal K}=2.1437$ is marked. The time differences in \eqref{eq:Sms} are first order, so the discrete
equations are first order in $\Delta t$. Table~\ref{tab:convergence} lists
the maximum relative deviation of $a$ and the residual of \eqref{eq:discRay}
for several step sizes. At $\tau=6$ the deviation is $3\%$ for
$\Delta t=0.05$ and $1.6\%$ for $\Delta t=0.025$.

\begin{table}[t]
\centering
\caption{Convergence of the discrete evolution on the 600-cell lattice. The
second column is the maximum relative deviation of $a(\tau)$ from the
continuum closed solution for $0<\tau\le6$. The third column is the maximum
relative residual of the discrete Raychaudhuri equation \eqref{eq:discRay}.}
\label{tab:convergence}
\begin{tabular}{ccc}
\hline
$\Delta t$ & $\max|\Delta a|/a$ & rel. residual of \eqref{eq:discRay} \\
\hline
$0.2$    & $1.1\times10^{-1}$ & $1.6\times10^{-1}$ \\
$0.1$    & $6.0\times10^{-2}$ & $8.6\times10^{-2}$ \\
$0.05$   & $3.1\times10^{-2}$ & $4.4\times10^{-2}$ \\
$0.025$  & $1.6\times10^{-2}$ & $2.2\times10^{-2}$ \\
$0.0125$ & $8.1\times10^{-3}$ & $1.1\times10^{-2}$ \\
\hline
\end{tabular}
\end{table}

\begin{figure}[t]
  \centering
  \includegraphics[width=\textwidth]{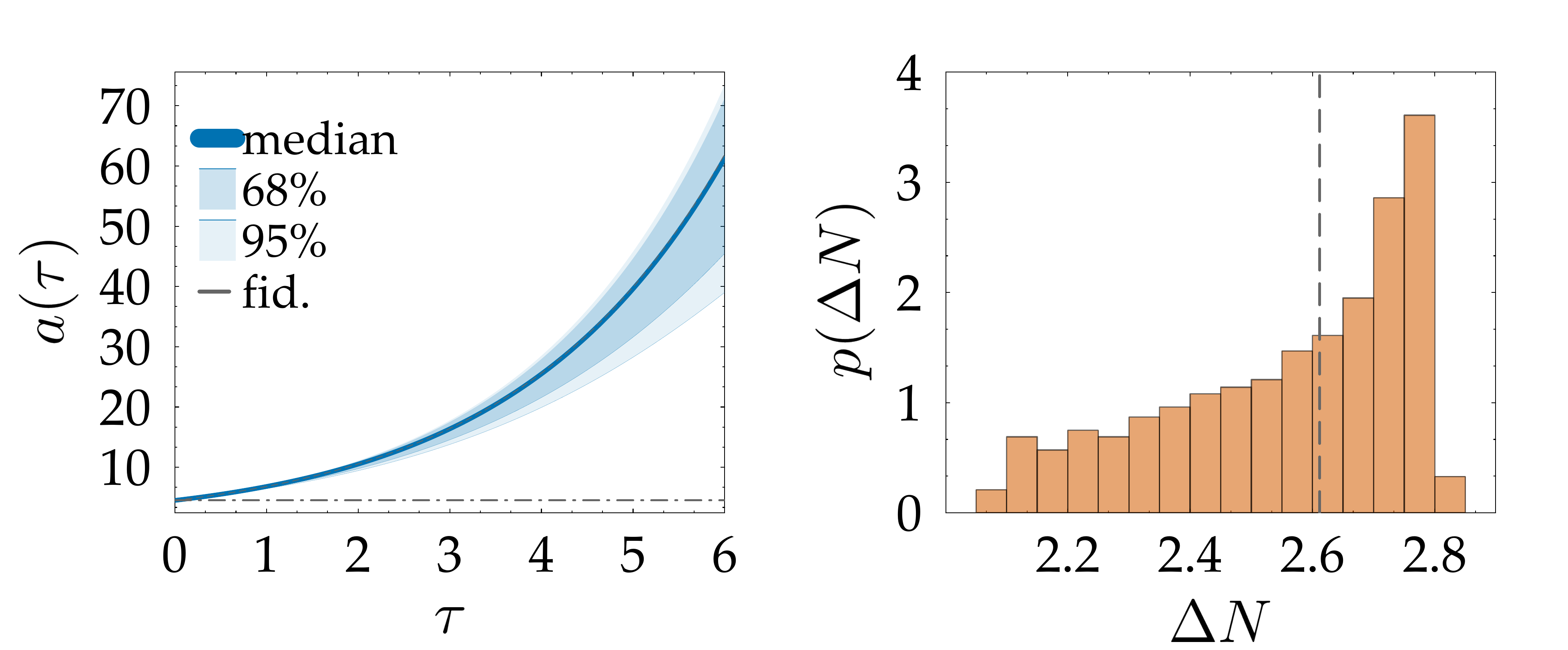}
  \caption{Monte Carlo ensemble of $2000$ initial conditions evolved with the
  discrete equations on the 600-cell lattice at $\Delta t=0.025$. The initial
  scalaron is drawn uniformly from $[2,4]$ and the axion charge from
  $[0,0.03]$, with $a(0)=4.5$. Left, median (solid), 16th/84th (dark band),
  and 5th/95th (light band) percentiles of $a(\tau)$; the dashed curve is the
  fiducial trajectory with $\phi(0)=3$ and $Q=0.02$. Right, probability
  density of $\Delta N=\ln[a(6)/a(0)]$, with the median marked by the dashed
  vertical line.}
  \label{fig:mc}
\end{figure}

To go beyond a single trajectory, we draw $2000$ initial conditions with
$\phi(0)$ uniform in $[2,4]$ and $Q$ uniform in $[0,0.03]$, fix $a(0)=4.5$,
and evolve each sample with the discrete equations at $\Delta t=0.025$. This
is a parameter scan, not a Bayesian inference. The uniform ranges are chosen
so that every sample has a bounce and the axion remains subdominant at the
bounce, and the spread of $\Delta N$ therefore measures the sensitivity to the
initial conditions rather than a posterior.
Figure~\ref{fig:mc} shows the resulting median and percentile bands of
$a(\tau)$ and the distribution of the number of e-folds
$\Delta N=\ln[a(6)/a(0)]$. The median is $\Delta N=2.61$, with $68\%$ of the
sample between $2.31$ and $2.76$. The spread is controlled mainly by the
initial value of the scalaron, which sets the slope of the potential, while
the axion charge has a subdominant effect at these values. The correlation
coefficient between $\phi(0)$ and $\Delta N$ is $0.96$, while $Q$ is
essentially uncorrelated with $\Delta N$.

\section{Conclusions}\label{sec:conclusion}

We have shown that, in the special case of metric-affine $f(R)$ gravity, the
theory can be formulated in
Regge calculus without ever defining torsion geometrically. Three ingredients
are essential. First, the axial torsion of the metric-affine $f(R)$ theory is
carried by a Kalb--Ramond two-form, while the trace part of the contorsion is
absorbed into the scalaron and the tensor part is eliminated algebraically.
Second, the torsion sector of the Einstein-frame action is a single perfect
square, $-\tfrac12\int|\mathcal H|^2$ with
$\mathcal H=\dd Z-2\,\dd w\wedge Z$, and the redefinition $Z=\Omega^2Z'$
linearizes it into a standard dilaton--Kalb--Ramond coupling. Third, in four
dimensions the two-form is dual to a scalar, so that the entire torsion
sector reduces to a single axion on the dual vertices, with the axial torsion
vector equal to the axion gradient.

The resulting discrete theory is pure Regge gravity coupled to two scalar
fields, with the explicit lattice action \eqref{eq:Slatt}. It requires
neither a $p\ge2$ discrete exterior calculus nor a geometric torsion, thereby
sidestepping the obstructions met by other approaches. We derived the complete set of discrete equations of
motion, namely the axion current conservation, the scalaron Klein--Gordon
equation, and the edge equation in which the deficit-angle variation is
removed by the Sch\"afli identity, and we verified that the edge equation
reduces to the Einstein equation in the continuum limit. On a homogeneous
lattice the equations reduce to a discrete Friedmann--Raychaudhuri system.
For a regular triangulation of $\mathbb S^3$ the kinetic coefficient is
$3V_0$ and the curvature coefficient is $N_1\theta_{sp}/2$, with the single
curvature number
$\mathcal K=6\sqrt2\,(2\pi-n_3\arccos\frac13)/n_3$ depending only on the
number of tetrahedra per edge. The discrete equations reproduce the
continuum closed solution at first order in $\Delta t$.

Along the way, we also showed that a generic local conformal transformation maps
a Regge metric out of the Regge class and that the deficit angles are not
conformally invariant. Consequently exact conformal dualities cannot hold
between Regge-type lattice theories, and the conformal mode cannot be
represented within Regge variables. This also explains why the lattice scalar
kinetic term necessarily couples simplices sharing only a hinge.

We emphasise that this  work is not a general solution of the problem of Regge calculus with torsion.
It is a rather special route that works because the torsion of the metric-affine
$f(R)$ theory is non-dynamical. The tensor part vanishes on shell, the trace
is slaved to the scalaron, and only the axial part survives, which is dual to
a single scalar. For a general Einstein--Cartan theory, or for any theory in
which the tensor part propagates, the obstructions discussed in the
introduction remain. The construction should therefore be read as a proof of
principle, a very special case that avoids the obstructions, not as a complete
framework. The lattice derivative is second order only for regular
simplices, so the discretization is also restricted to the regular
triangulations used here.

Several directions deserve further study. First, the path integral over the
scale factor and the scalaron. The Euclidean measure is obstructed by the
conformal-factor problem, since the scale-factor kinetic term in the
Lorentzian action has the wrong sign. A controlled Monte Carlo requires a
contour rotation or a Picard--Lefschetz thimble, and we leave it for future
work. The Jacobian of the field-dependent redefinition $Z=\Omega^2Z'$ also
remains to be fixed. Second, one could extend the discrete equations to
inhomogeneous and anisotropic lattices. Third, one may ask whether the
conformal non-invariance can be evaded by enlarging the lattice to include
non-flat simplices.
\section*{Acknowledgements} 
The author would like to thank Ruijue Yan for helpful discussions.
X.G. is supported by Yancheng Institute of Technology (xjr2024030). 
\appendix
\section{A single tetrahedron example for conformal non-invariance}\label{AnnA}
Two properties of the continuum transformation are in tension on the lattice.
The map $g_{ab}\mapsto\Omega^2g_{ab}$ rescales lengths by $\Omega$ and preserves angles.
A Regge metric carries no independent angle data, since the angles of each
simplex are fixed by its edge lengths. The natural discrete version of the
length rescaling is $l_{ij}\to\tfrac12(\Omega_i+\Omega_j)l_{ij}$, with
$\Omega_i$ the value of $\Omega$ at the vertex $P_i$. This is a pure length
rescaling, and it does not preserve angles, because a vertex-dependent factor
changes the ratios of the edge lengths of a simplex and hence its dihedral
angles. Conversely, if one demands that the transformation preserve all
angles, the only allowed factor is a global constant, since a local
$\Omega(x)$ would have to be a similarity on every simplex and could not vary
from simplex to simplex without changing the shapes. There is therefore no
nontrivial local conformal transformation that stays within the Regge class.
The example below makes this explicit.

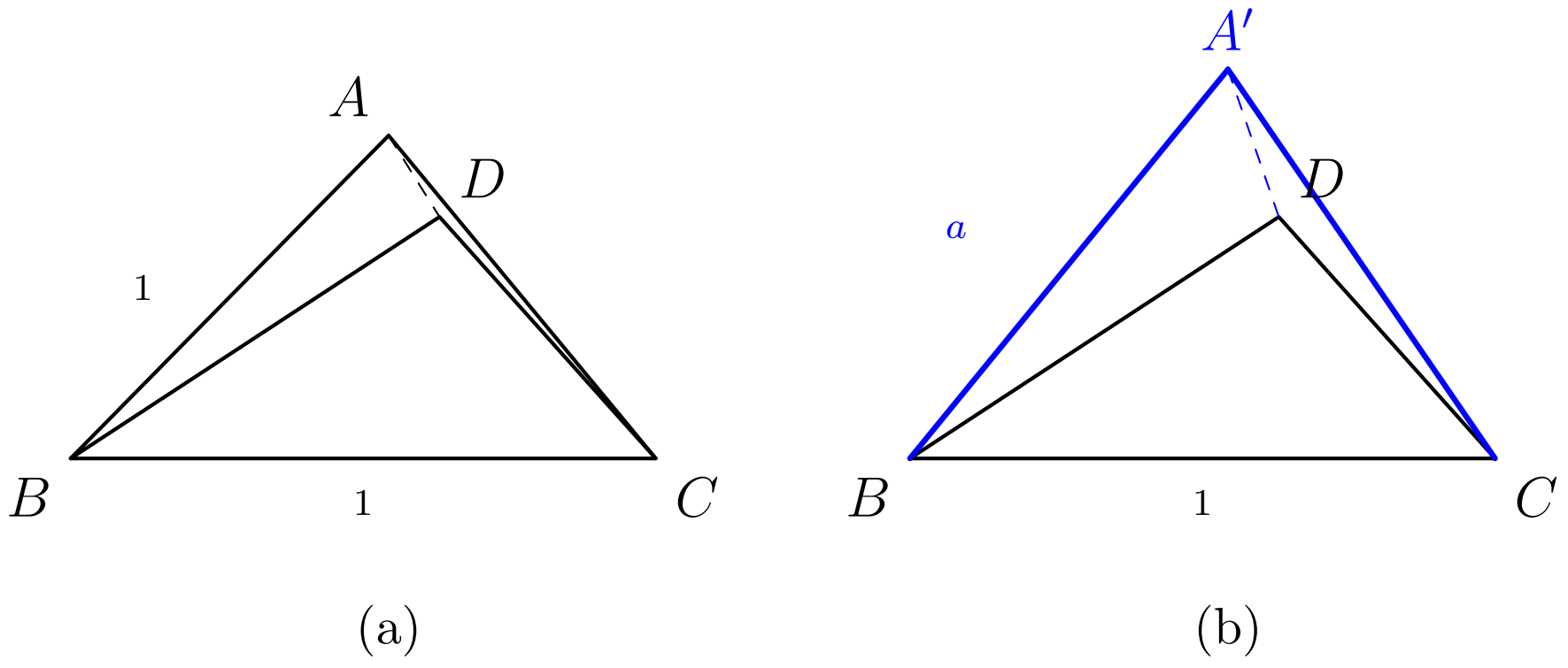
\begin{figure}[t]
\centering
\begin{tikzpicture}[scale=1.9,line join=round,line cap=round]
  \coordinate (B) at (-1.15,-0.45);
  \coordinate (C) at (1.15,-0.45);
  \coordinate (D) at (0.3,0.5);
  \coordinate (A) at (0.1,0.82);
  \draw[thick] (B)--(C)--(D)--cycle;
  \draw[thick] (A)--(B);
  \draw[thick] (A)--(C);
  \draw[dashed] (A)--(D);
  \node[below left] at (B) {$B$};
  \node[below right] at (C) {$C$};
  \node[above right] at (D) {$D$};
  \node[above left] at (A) {$A$};
  \node[below,font=\scriptsize] at (0,-0.5) {$1$};
  \node[left,font=\scriptsize] at (-0.75,0.22) {$1$};
  \node[below,font=\small] at (0.1,-0.95) {(a)};
  \begin{scope}[xshift=3.3cm]
    \coordinate (B) at (-1.15,-0.45);
    \coordinate (C) at (1.15,-0.45);
    \coordinate (D) at (0.3,0.5);
    \coordinate (A) at (0.1,1.08);
    \draw[thick] (B)--(C)--(D)--cycle;
    \draw[very thick,blue] (A)--(B);
    \draw[very thick,blue] (A)--(C);
    \draw[dashed,blue] (A)--(D);
    \node[below left] at (B) {$B$};
    \node[below right] at (C) {$C$};
    \node[above right] at (D) {$D$};
    \node[above,blue] at (A) {$A'$};
    \node[below,font=\scriptsize] at (0,-0.5) {$1$};
    \node[left,blue,font=\scriptsize] at (-0.85,0.45) {$a$};
    \node[below,font=\small] at (0.1,-0.95) {(b)};
  \end{scope}
\end{tikzpicture}
\caption{Conformally rescaled tetrahedron. (a) The regular tetrahedron with
unit edges. (b) The base $BCD$ is held fixed while the three edges from the
apex are rescaled to $a=1+\epsilon/2$, moving the apex to $A'$. The angle at
$B$ in each isosceles face is $\beta=\arccos(1/2a)$, so the deficit angle at
$B$ is $\theta_B=2\pi-[\pi/3+2\beta]$.}
\label{fig:tetra}
\end{figure}

The mechanism is visible in a single tetrahedron (Fig.~\ref{fig:tetra}). Let $ABCD$ be a regular tetrahedron with all edges of length $1$, and rescale
the edge lengths by the vertex factors $\Omega_A=1+\epsilon$ and
$\Omega_B=\Omega_C=\Omega_D=1$, that is, $l_{ij}\to\bar\Omega_{ij}l_{ij}$ with
$\bar\Omega_{ij}=\tfrac12(\Omega_i+\Omega_j)$. For the three edges out of $A$
this gives $1+\epsilon/2$, while for the three edges of the base triangle
$BCD$ this gives $1$. Hence the base triangle remains equilateral of side
$1$, and the three edges emanating from $A$ acquire length $a=1+\epsilon/2$.

The faces incident on $B$ are $BCD$ (equilateral) and the two isosceles
faces $ABC$ and $ABD$ with sides $(a,a,1)$. The angle at $B$ in an isosceles
triangle follows from the law of cosines,
\begin{equation}
\cos\angle_B=\frac{a^2+1-a^2}{2a}=\frac{1}{2a},
\end{equation}
and the angle at $B$ in the equilateral face is $\pi/3$, so
\begin{equation}
\theta_B(a)=2\pi-\Big[2\arccos\frac{1}{2a}+\frac{\pi}{3}\Big].
\label{eq:thetaB}
\end{equation}
Differentiating and evaluating at $a=1$ gives
$\frac{\dd\theta_B}{\dd a}|_{a=1}=-\frac{2}{\sqrt3}$. Since
$a-1=\epsilon/2$,
\begin{equation}
\Delta\theta_B=-\frac{\epsilon}{\sqrt3}+\mathcal O(\epsilon^2).
\label{eq:deficitchange}
\end{equation}
Thus individual deficit angles are not conformally invariant. The total
deficit of a triangulated $\mathbb S^2$ is $2\pi\chi=4\pi$, protected by the
Gauss--Bonnet theorem. The three base vertices are equivalent, so
$\Delta\theta_A=-3\Delta\theta_B+\mathcal O(\epsilon^2)$, that is,
$\Delta\theta_A=\sqrt3\,\epsilon+\mathcal O(\epsilon^2)$.

Comparing the two discretizations of $\int\dd^dx\sqrt{\tilde g}\,\tilde R$, (A) on the rescaled edge lengths and (B)
from the transformed curvature,
\begin{equation}
\text{(A)}-\text{(B)}=\mathcal O(l^2).
\label{eq:residual}
\end{equation}
For (B) this follows
from \eqref{eq:confR} and $\sqrt{\tilde g}=\Omega^4\sqrt g$, discretizing the
right-hand side with \eqref{eq:derivcont}. For a smooth conformal factor one
has $\epsilon\sim l\,\hat\partial w$, so the individual deficit angles change
by $O(\epsilon)$ [Eq.~\eqref{eq:deficitchange}] and the change of the Regge
action is a discretization effect of the same order as \eqref{eq:residual}.
The two discretizations therefore agree only up to the discretization error,
and the equality cannot hold simplex by simplex.

The contradiction is now apparent. The rescaled tetrahedron is an admissible
Regge metric, but its deficit angles differ from those of the original by
$\mathcal O(\epsilon)$. A conformal transformation preserves angles and would
leave them unchanged. The edge-rescaled metric is therefore not the conformal
image of the original metric. A true conformal image is generically not
piecewise flat, and the two descriptions agree only up to the discretization
error, as \eqref{eq:residual} shows.

\section{Homogeneous reduction and the regular triangulation}\label{app:frw}
With \begin{equation}ds^2=-N^2dt^2+\ell^2\gamma_{ij}dx^idx^j,\end{equation} where $\gamma$ has unit
curvature and $\int d^3x\sqrt\gamma=V_0$, one has
\begin{equation}h_{ij}=\ell^2\gamma_{ij},\quad K_{ij}=\frac{\ell\dot\ell}{N}\gamma_{ij},\quad
K=\frac{3\dot\ell}{N\ell},\quad K_{ij}K^{ij}=\frac{3\dot\ell^2}{N^2\ell^2}.\end{equation} Hence
$K_{ij}K^{ij}-K^2=-\frac{6\dot\ell^2}{N^2\ell^2}$, ${}^{(3)}R=\frac{6k}{\ell^2}$,
and $\sqrt{-g}=N\ell^3\sqrt\gamma$. Using
$\int\!\sqrt{-g}\,R=\int\!\sqrt{-g}\,({}^{(3)}R+K_{ij}K^{ij}-K^2)$, we have
\begin{equation}
\frac{1}{2\kappa^2}\int\!\sqrt{-g}\,R
=\frac{V_0}{\kappa^2}\int\!dt\Big[-\frac{3\ell\dot\ell^2}{N}+3Nk\ell\Big].
\label{eq:adm1}
\end{equation}

With $(\partial\phi)^2=-(\dot\phi/N)^2$, the scalar actions become
\begin{equation}
S_\phi=V_0\!\int\!dt\Big[\frac{\ell^3\dot\phi^2}{2N}-N\ell^3W\Big],
\qquad
S_a=V_0\!\int\!dt\,\frac{\Omega^{-4}\ell^3\dot a^2}{2N}.
\label{eq:matterms}
\end{equation}

Replacing $\dot q\to\Delta q_n/\Delta t$ and $\int dt\to\sum_n\Delta t$ gives the action
\eqref{eq:Sms},
\begin{equation}
S_{\rm ms}=V_0\sum_n\Big[
-\frac{3\ell_n(\Delta\ell_n)^2}{\kappa^2N_n\Delta t}
+\frac{3N_nk\ell_n\Delta t}{\kappa^2}
+\frac{\ell_n^3(\Delta\phi_n)^2}{2N_n\Delta t}
-N_n\Delta t\,\ell_n^3W_n
+\frac{\Omega^{-4}\ell_n^3(\Delta a_n)^2}{2N_n\Delta t}\Big].
\end{equation}
Let us write $S_{\rm ms}=V_0\sum_nL_n$.
Varying the terms of $L_n$,
\begin{equation}
\frac{\partial L_n}{\partial N_n}
=\frac{3\ell_n(\Delta\ell_n)^2}{\kappa^2N_n^2\Delta t}
+\frac{3k\ell_n\Delta t}{\kappa^2}
-\frac{\ell_n^3(\Delta\phi_n)^2}{2N_n^2\Delta t}
-\Delta t\,\ell_n^3W_n
-\frac{\Omega^{-4}\ell_n^3(\Delta a_n)^2}{2N_n^2\Delta t},
\end{equation}
so $\partial L_n/\partial N_n=0$ gives
\begin{equation}
3H_n^2=\kappa^2\Big[\tfrac12\dot\phi_n^2+W_n
+\tfrac12\Omega^{-4}\dot a_n^2\Big]-\frac{3k}{\ell_n^2}.
\end{equation}
Since $\Delta\ell_n$ and $\Delta\ell_{n-1}$ both contain $\ell_n$,
\begin{equation}
\frac{\partial L_n}{\partial\ell_n}
=-\frac{3}{\kappa^2}\frac{(\Delta\ell_n)^2-2\ell_n\Delta\ell_n}{N_n\Delta t}
+\frac{3kN_n\Delta t}{\kappa^2}
+\frac{3\ell_n^2(\Delta\phi_n)^2}{2N_n\Delta t}
-3N_n\Delta t\,\ell_n^2W_n
+\frac{3\Omega^{-4}\ell_n^2(\Delta a_n)^2}{2N_n\Delta t},
\end{equation}
\begin{equation}
\frac{\partial L_{n-1}}{\partial\ell_n}
=-\frac{6}{\kappa^2}\frac{\ell_{n-1}\Delta\ell_{n-1}}{N_{n-1}\Delta t}.
\end{equation}
Setting the sum to zero and multiplying by $-\kappa^2\Delta t/3$ yields
\eqref{eq:discRay}.
For the scalaron,
\begin{equation}
\frac{\partial L_n}{\partial\phi_n}
=-\frac{\ell_n^3\Delta\phi_n}{N_n\Delta t}
-N_n\Delta t\,\ell_n^3W_n'
+\tfrac12\frac{\dd\Omega^{-4}}{\dd\phi}\frac{\ell_n^3(\Delta a_n)^2}{N_n\Delta t},
\qquad
\frac{\partial L_{n-1}}{\partial\phi_n}
=+\frac{\ell_{n-1}^3\Delta\phi_{n-1}}{N_{n-1}\Delta t},
\end{equation}
which, with $\frac{\dd\Omega^{-4}}{\dd\phi}=-\frac{4\kappa}{\sqrt6}\Omega^{-4}$,
gives \eqref{eq:discKG}. For the axion, $a_n$ enters only through the axion
kinetic terms, which gives directly \eqref{eq:discaxion}.

Expanding
$\Delta q_n=\dot q\,N_n\Delta t+\tfrac12\ddot q\,N_n^2\Delta t^2
+O(\Delta t^3)$, the $O(\Delta t)$ pieces cancel and \eqref{eq:discRay}
becomes
\begin{equation}
-\big(\dot\ell^2+2\ell\ddot\ell\big)
=\kappa^2\ell^2\Big[\tfrac12\dot\phi^2+\tfrac12\Omega^{-4}\dot a^2-W\Big]+k,
\end{equation}
that is,
$2\ddot\ell/\ell+(\dot\ell/\ell)^2=-\kappa^2 p-k/\ell^2$, which is equivalent to
$2\dot H+3H^2=-\kappa^2 p-k/\ell^2$. Together with the lapse equation this is
the Friedmann--Raychaudhuri system \eqref{eq:contFried}.

The 5-, 16-, and 600-cell triangulations used below are those of the
Collins-Williams Regge cosmology \cite{Collins1973,Liu2016}. Consider a
regular tetrahedron of edge length $2\sqrt2$ with vertices
$(1,1,1)$, $(1,-1,-1)$, $(-1,1,-1)$, and $(-1,-1,1)$. Its facet normals are
\begin{equation}
n_{ABC}=(4,4,-4),\qquad n_{ABD}=(-4,4,-4),
\end{equation}
so that
\begin{equation}
\cos\theta_\Delta=\frac{n_{ABC}\cdot n_{ABD}}{|n|\,|n|}
=\frac{16}{48}=\frac13,\qquad\theta_\Delta=\arccos\tfrac13,
\end{equation}
and the volume is $8/3$. Rescaling to unit edge gives
$V_\Delta=1/(6\sqrt2)$, so that
\begin{equation}
V_0=N_3V_\Delta=\frac{N_3}{6\sqrt2},\qquad N_1=\frac{6N_3}{n_3},
\qquad \theta_{sp}=2\pi-n_3\arccos\tfrac13 .
\end{equation}
The induced spatial curvature is exact for the equilateral slice,
\begin{equation}\int\sqrt h\,{}^{(3)}R=N_1\ell\theta_{sp},\end{equation} and matching $6kV_0\ell$ to
$N_1\ell\theta_{sp}$ gives \eqref{eq:Knum}. The extrinsic-curvature part is
$-3V_0\ell\dot\ell^2/N$, that is, the kinetic coefficient $3V_0$. This yields
\eqref{eq:Smsreg} and \eqref{eq:discFriedreg} with the numerical values quoted in
Sec.~\ref{sec:frw}.

\section{The prism triangulation and the mixed-hinge sum}\label{app:prism}

This appendix substantiates the claim, made in Sec.~\ref{sec:frw}, that the
homogeneous reduction of the Regge term involves a well-defined but
constrained finite sum over mixed hinges, and it records an obstruction to
homogeneity.

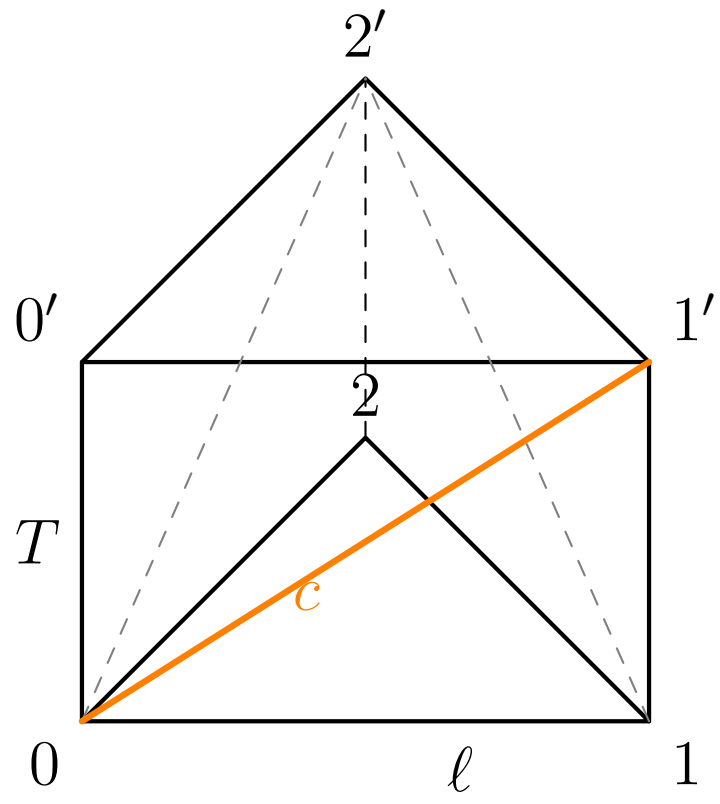
\begin{figure}[t]
\centering
\begin{tikzpicture}[scale=1.25,line join=round,line cap=round]
  \coordinate (0) at (-1.5,-0.7);
  \coordinate (1) at (1.5,-0.7);
  \coordinate (2) at (0,0.8);
  \coordinate (0') at (-1.5,1.2);
  \coordinate (1') at (1.5,1.2);
  \coordinate (2') at (0,2.7);
  \draw[thick] (0)--(1)--(2)--cycle;
  \draw[thick] (0')--(1')--(2')--cycle;
  \draw[thick] (0)--(0');
  \draw[thick] (1)--(1');
  \draw[dashed] (2)--(2');
  \draw[gray,dashed] (0)--(1');
  \draw[gray,dashed] (0)--(2');
  \draw[gray,dashed] (1)--(2');
  \draw[very thick,orange] (0)--(1');
  \node[below left] at (0) {$0$};
  \node[below right] at (1) {$1$};
  \node[above] at (2) {$2$};
  \node[above left] at (0') {$0'$};
  \node[above right] at (1') {$1'$};
  \node[above] at (2') {$2'$};
  \node[below] at (0.5,-0.72) {$\ell$};
  \node[left] at (-1.5,0.25) {$T$};
  \node[below right,orange] at (-0.5,0.15) {$c$};
\end{tikzpicture}
\caption{Two-dimensional analogue of the staircase triangulation. The
triangular prism $\Delta^2\times I$ is split into the three tetrahedra
$S_0=(0,0'1'2')$, $S_1=(0,1,1'2')$, and $S_2=(0,1,2,2')$. The spatial and
timelike edges are $\ell$ and $T$, the cross edges $ij'$ are dashed, and the
highlighted edge $01'$ is one staircase step with length $c$. In four
dimensions the same construction splits $\Delta^3\times I$ into the four
$4$-simplices $S_0$--$S_3$ of Eq.~\eqref{eq:staircase}.}
\label{fig:staircase}
\end{figure}

Take two consecutive slices, each carrying a regular triangulation of
$\mathbb S^3$ with vertices $v_i$ and $v_{i'}$. For each spatial tetrahedron
$(0123)$ the product $\Delta^3\times I$ is triangulated into four
$4$-simplices by the staircase \cite{DeLoera2010} (Fig.~\ref{fig:staircase})
\begin{equation}
S_0=(0,0'1'2'3'),\quad S_1=(0,1,1'2'3'),\quad
S_2=(0,1,2,2'3'),\quad S_3=(0,1,2,3,3'),
\label{eq:staircase}
\end{equation}
with $S_k$ containing the bottom vertices $v_0,\dots,v_k$ and the top
vertices $v_k',\dots,v_3'$. Besides the spatial edges $\ell$ and the
timelike edges $T=N\Delta t$, the staircase introduces \emph{cross edges}
$ij'$ with $i<j$, whose lengths are not determined by $\ell$ and $T$. For a
homogeneous ansatz all cross edges are equal, $ij'=c$, but $c$ is an
independent edge variable, fixed either by its own equation of motion
$\partial S/\partial c=0$ or by a smoothness prescription. For the straight
prism embedding one has $c^2=\ell^2+T^2$. We use this value below
and set $\ell=1$.

For a $d$-simplex the dihedral angle $\theta_\tau$ at the codimension-two
face $\tau$ is
\begin{equation}
\sin\theta_\tau=\frac{d\;V_d\;V_{d-2}(\tau)}{(d-1)\,V_{d-1}(F_1)\,V_{d-1}(F_2)},
\label{eq:volformula}
\end{equation}
where $F_{1,2}$ are the two facets sharing $\tau$. For $d=4$,
\begin{equation}
\sin\theta_\tau=\frac{4\,V_4\,A_\tau}{3\,V_3(F_1)\,V_3(F_2)} .
\label{eq:volformula4}
\end{equation}
The volumes are given by the Cayley--Menger determinants
\begin{equation}
V_n^2=(-1)^{n+1}\frac{\det M_n}{2^n(n!)^2},\qquad
M_n=\begin{pmatrix}0&1&\cdots&1\\ 1&&&\\ \vdots&&d_{ij}^2&\\ 1&&&\end{pmatrix},
\label{eq:CM}
\end{equation}
and the sign of $\theta$ is fixed by the outward facet normals,
$\cos\theta_\tau=-\hat n_1\!\cdot\!\hat n_2$. Formula \eqref{eq:volformula}
can be verified by the regular simplices. For $d=2$ it gives
$\theta=\pi/3$. For $d=3$ it gives $\theta=\arccos(1/3)$. For $d=4$ it gives
$\theta=\arccos(1/4)$, for which \eqref{eq:volformula4} returns
$\sqrt{15}/4$. 

By their side lengths under the symmetries of the
staircase, the mixed hinges fall into two families,
(I) $(\ell,c,T)$ and (II) $(\ell,c,c)$:
\paragraph{Family (I).} For $\tau=(0,1,1')$ the hinge is right-angled with
legs $\ell=1$ and $T$, so $A_\tau=T/2$. The two adjacent facets
$F_a=(0,1,1',2')$ and $F_b=(0,1,1',3')$ have equal volume
$V_3=\sqrt3\,T/12$, and the $4$-simplex has $V_4=T/(24\sqrt2)$. Equation
\eqref{eq:volformula4} gives
\begin{equation}
\sin\theta_{\rm I}=\frac{2\sqrt2}{3}.
\end{equation}
The outward normals give $\cos\theta_{\rm I}=1/3$, hence
\begin{equation}
\theta_{\rm I}=\arccos\tfrac13,\qquad
\delta_{\rm I}=2\pi-n_3\arccos\tfrac13 ,
\label{eq:thetaI}
\end{equation}
with $n_3$ the number of adjacent $4$-simplices. This angle is independent
of $T$.

\paragraph{Family (II).} For $\tau=(0,1,2')$ the hinge area is
$A_\tau=\tfrac12\sqrt{T^2+\tfrac34}$, and the two adjacent facets have
$V_3=\sqrt3\,T/12$ and $\tfrac16\sqrt{T^2+\tfrac12}$, with the same $V_4$.
Equation \eqref{eq:volformula4} gives
\begin{equation}
\sin\theta_{\rm II}
=\frac{\sqrt{4T^2+3}}{\sqrt3\,\sqrt{2T^2+1}}.
\label{eq:thetaII}
\end{equation}
The outward normals give
$\cos\theta_{\rm II}=T\sqrt{2/3}/\sqrt{1+2T^2}$, so
$\theta_{\rm II}\in(0,\pi/2)$. Four $4$-simplices meet at this hinge, hence
$\delta_{\rm II}=2\pi-4\theta_{\rm II}$, and $\delta_{\rm II}\to0$ as
$T\to0$.

 Since each mixed hinge contributes
$A_\tau\sin\delta_\tau\propto T$, the mixed contribution is of order $T$
and, for a smooth configuration, it reproduces the extrinsic-curvature term
of \eqref{eq:Sms}, that is, the kinetic coefficient $3V_0$ of
Eq.~\eqref{eq:discFriedreg}, up to $\mathcal O(l^2)$ corrections.

The staircase triangulation depends on an ordering $0<1<2<3$ and is not
invariant under the spatial symmetry group, which changes which faces exist.
Consequently the $n_3$ tetrahedra around an edge are not all equivalent and
the local geometry is not vertex-transitive, so that the homogeneous ansatz
of Sec.~\ref{sec:frw} cannot be realized on the bare staircase complex. Only
aligned corner faces $(a,b,b')$ and $(a,a',c')$ occur, not $(a,b,a')$. Three
remedies would restore homogeneity. One may use a triangulation of $\Delta^3\times I$
independent of a vertex ordering, average over all orderings, or use a coned
construction in which all $4$-simplices are congruent, so that the mixed
hinges form a single orbit with deficit angle \eqref{eq:thetaI}. The
obstruction is confined to the product and staircase constructions and does
not affect the topological or continuum results.


\end{document}